\documentclass[reprint,showkeys,sort&compress]{revtex4-1}

\usepackage{graphicx}
\usepackage{subfig}
\usepackage{wasysym}
\usepackage{enumerate}
\usepackage{hyperref}

\hypersetup{
    colorlinks=true,%
    citecolor=blue,%
    filecolor=blue,%
    linkcolor=blue,%
    urlcolor=blue
}
\begin{document}
\title{Are Epileptic Seizures Quakes of the Brain? An Approach by Means of Nonextensive Tsallis Statistics}

%
%
%
%
%

\author{K. Eftaxias}
\affiliation{Department of Physics, Section of Solid State Physics, University of Athens, Panepistimiopolis, GR 15784, Zografos, Athens, Greece; email: ceftax@phys.uoa.gr}

\author{G. Minadakis}
\affiliation{Department of Electronic and Computer Engineering, Brunel University, Kingston Lane, Uxbridge, Middlesex, UB8 3PH, U.K.; email: george.minadakis@brunel.ac.uk}

\author{L. Athanasopoulou}
\affiliation{Department of Physics, Section of Solid State Physics, University of Athens, Panepistimiopolis, GR-15784, Zografos, Athens, Greece; email: labrini@gmail.com}

\author{M. Kalimeri}
\affiliation{Department of Physics, Section of Solid State Physics, University of Athens, Panepistimiopolis, GR-15784, Zografos, Athens, Greece; email: mkalime@gmail.com}

\author{S. M. Potirakis}
\affiliation{Department of Electronics, Technological Education Institute of Piraeus, 250 Thivon \& P. Ralli, GR-12244, Aigaleo, Athens, Greece; email: spoti@teipir.gr}

\author{G. Balasis}
\affiliation{Institute for Space Applications and Remote Sensing, National Observatory of Athens, Metaxa and Vas. Pavlou St., Penteli, 15236 Athens, Greece, email: gbalasis@space.noa.gr}




\begin{abstract}
The field of study of complex systems holds that the dynamics of complex systems are founded on universal principles that may used to describe a great variety of scientific and technological approaches of different types of natural, artificial, and social systems. Authors have suggested that earthquake dynamics and neurodynamics can be analyzed within similar mathematical frameworks, a claim further supported by recent evidence. The purpose of this paper is to suggest a shift in emphasis from the large to the small in the search for a dynamical analogy between seizure and earthquake. Our analyses focus on a single epileptic seizure generation and the activation of a single fault (earthquake) and not on the statistics of sequences of different seizures and earthquakes. A central property of the epileptic seizure / earthquake generation is the occurrence of coherent large-scale collective behaviour with very rich structure, resulting from repeated nonlinear interactions among the constituents of the system, respectively firing neurons and opening cracks. Consequently, in this paper, we apply the Tsallis nonextensive statistical mechanics as it proves an appropriate framework in order to investigate universal principles of their generation. For completeness reasons we also use entropic measures as well as tools from information theory. The obtained results seems to support the claim that epileptic seizures can be considered as ``quakes on the brain''.
\end{abstract}

\maketitle
\section{Introduction}

In the last 15 years, the study of complex systems has emerged as a recognized field in its own right, although, a good definition of what a complex system is has proven elusive. The very concept of complexity is nowadays frequently used yet poorly defined -at least quantitatively speaking- and its study has embraced a great variety of scientific and technological approaches of all types of natural, artificial, and social systems. When one considers a phenomenon or a thing that is ``complex'', one generally associates it with something that is ``hard to separate, analyze or solve''. Instead, we refer to ``a complex system'' as one whose phenomenological laws, which describe the global behaviour of the system, are not necessarily directly related to the ``microscopic'' laws that regulate the evolution of its elementary parts. In other words, ``complexity'' is the emergence of a non-trivial behaviour due to the interactions of the subunits that form the system itself. The statistical features of complex systems are generally not dependent on the details of the interacting subunits that form the system. Another relevant ingredient of a complex system is that topological disorder within the system will generally introduce new, surprising effects, different than those one would expect from the simple ``microscopic'' evolution rules.

The field of study of complex systems holds that the dynamics of complex systems are founded on universal principles that may used to describe disparate problems ranging from particle physics to economies of societies \cite{Bar-Yam1997}. This is a basic reason for our interest in complexity \cite{Stanley1999,Stanley2000a,Sornette2002,Vicsek2001,Vicsek2002}. Empirical evidence has been mounting that supports the possibility that a number of systems under study in disciplines as diverse as physics, biology, engineering, and economics may have certain quantitative features that are intriguingly similar. Picoli et. al. \cite{Picoli2007} reported similarities between the dynamics of geomagnetic signals and heartbeat intervals. de Arcangelis et al. \cite{Arcangelis2006} presented evidence for universality in solar flare and earthquake occurrence. Kossobokov and Keilis-Borok \cite{Kossobokov2000} have explored similarities of multiple fracturing on a neutron star and on the Earth, including power-law energy distributions, clustering, and the symptoms of transition to a major rupture. Sornette and Helmstetter \cite{Sornette2002} have presented occurrence of finite-time singularities in epidemic models of rupture, earthquakes, and starquakes. Abe and Suzuki \cite{Abe2004} have shown that Internet shares  common scale-invariant features with earthquakes, in its temporal behaviours. Fukuda et al. \cite{Fukuda2003} have reported similarities between communication dynamics in the Internet and the automatic nervous system. Peters et al. \cite{Peters2002} have shown that the rain events are analogous to a variety of nonequilibrium relaxation processes in Nature such as earthquakes and avalanches. A corollary is that transferring ideas, methods and insights from investigations in hitherto disparate areas will cross-fertilize and lead to important new results.

Epileptic seizures and pre-seismic electromagnetic (EM) emissions rooted in the activation of a single fault are complex phenomena, which have highly intricate cluster and hierarchical structures, spatial and temporal correlation with feedback, self-organization and connection diversity. Authors have suggested that earthquake dynamics and neurodynamics can be analyzed within similar mathematical frameworks \cite{Herz1995,Rundle2002}. Characteristically, driven systems of interconnected blocks with stick-slip friction capture the main features of earthquake process. These models, in addition to simulating the aspects of earthquakes and frictional sliding, may also represent the dynamics of neurological networks \cite{Herz1995}. Hopfield \cite{Hopfield1994} proposed a model for a network of $N$ integrate-and-fire neurons. In this model, the dynamical equation of $k^{th}$ neuron, see equation 28 in \cite{Hopfield1994} is based on the Hodgekin-Huxley model for neurodynamics and represents the same kind of mean field limit that has been examined in connection with earthquakes (EQs) \cite{Rundle2002}.

Recently, Osorio et al. \cite{Osorio2010} in a pioneering work have shown that a dynamical analogy supported by five scale-free statistics (the Gutenberg-Richter distribution of event sizes, the distribution of intervals, the Omori laws, and the conditional waiting time until the next event) exists between seizures and earthquakes. More precisely, the authors performed the analysis using: (i) 81 917 earthquakes between 1984-2000 available in the Southern California Seismic Network catalogue; and (ii) 16.032 seizures from continuous multiday voltage recordings directly from the brains of 60 human subjects with mesial temporal and frontal lobe pharmacoresistant epilepsy undergoing surgical evaluation at the University of Kansas Medical Center between 1996 and 2000.

{\it It might be better though if the comparison between seizures and earthquakes is performed at the level of a single fault / seizure activation}. In this direction, the present work attempts to examine whether a unified theory may exist for the ways in which firing neurons / opening cracks organize themselves to produce a single epileptic seizure / earthquake. The presented investigation is developed in two stages. First, we examine the data in terms of multidisciplinary statistical analysis methods, aiming to discover common ``pathological'' symptoms of transition to a large seismic or epileptic shock. Then, we examine the existence of dynamical correspondences between seizure and earthquake generation by means of scale-free statistics, namely, a common hierarchical organization that results in power-law behaviour over a wide range of values of some parameter such as event energy or waiting time. We concentrate on the question whether the corresponding power-laws, if any, share the same exponent.

This paper is organized as follows. In the next section, we refer to the data collection. In Sec. 3 we introduce the essential concepts of nonextensive statistical mechanics. In Sec. 4, we briefly describe Tsallis entropy, T-entropy, approximate entropy, Block entropies, Fisher information and R/S analysis. In Sec. 4, we present the methods of statistical analysis which are applied to the data. The application of various measures of organization or information content in the epileptic and preseismic time series under study is presented in Sec. 5. In Sec. 6 we establish the hypothesis that a dynamical analogy can be found between epileptic seizure and activation of a single fault by means of Gutenberg-Richter law. In Sec. 7 we examine the existence of analogies in terms of nonextensivity. Following, dynamical analogy by means of waiting times is investigated in Sec. 8. In Sec. 9, we study the role of the coupling strength (or heterogeneity). The main results of the present work are discussed and summarized in Sec. 10.

\section{Data Collection}

{\it Electroencephalograms} (EEG) are brain signals provides us with information about the mean brain electrical activity, as measured at different sites of the head. EEGs not only provide insight concerning important characteristics of the brain activity but also yield clues regarding the underlying associated neural dynamics. The processing information by the brain is reflected in dynamical changes in this electrical activity.

{\it Electromagnetic seismograms} are EM signals which provide us with information about the fracture induced electromagnetic activity. Electromagnetic seismograms not only provide insight concerning important characteristics of the underlying fracture process but also yield clues regarding the associated fracture dynamics.

\subsection{Human EEGs}

The used data have been offered by Andrzejak et al. \cite{Andrzejak2001}. Two sets, denoted ``A'' and ``E'', respectively, each containing 100 single-channel EEG segments of $23.6 sec$ duration, were employed for this study. Set ``A'' is comprised of EEGs of healthy volunteers. Set ``E'' contain seizure activities. The segments fulfil the criterion of stationarity \cite{Andrzejak2001}. After 12 bit analog-to-digital conversion, the data were written continuously onto the disk of a data acquisition computer system at a sampling rate of 173.61 Hz. Band-pass filter settings were 0.53-40 Hz (12 dB/oct.).

\subsection{Rat EEG}

Adult Sprague-Dawley rats were used to study the epileptic seizures in EEG recordings \cite{Li2005,Kapiris2005}. The rats were anaesthetized with an i.p. injection of Nembutal (sodium pentobarbital, 65 mg/kg of body weight), and mounted in a stereotaxic apparatus. An electrode was placed in epidural space to record the EEG signals from temporal lobe. The animals were housed separately postoperatively with free access to food and water, allow 2-3 days to recover, and handled gently to familiarize them with the recording procedure. Each rat was initially anaesthetized with a dose of pentobarbital (60 mg/kg, i.p.), while constant body temperature was maintained (36.5-37.5$^{\circ}$C) with a piece of blanket. The degree of anaesthesia was assessed by continuously monitoring the EEG, and additional doses of anaesthetic were administered at the slightest change towards an awake pattern (i.e., an increase in the frequency and reduction in the amplitude of the EEG waves). Then, bicuculline i.p. injection was used to induce the rat epileptic seizures. EEG signals were recorded using an amplifier with band-pass filter setting of 0.5-100 Hz. The sampling rate was 200 Hz, and the analogue-to-digital conversion was performed at 12-bit resolution. The seizure onset time is determined by visual identification of a clear electrographic discharge, and then looking backwards in the record for the earliest EEG changes from baseline associated with the seizure. The earliest EEG change is selected as the seizure onset time. The interval between the seizure onset time and injection time are considered as the maximum prediction duration or extended pre-ictal phase.

\subsection{Pre-seismic EM emissions}

A question effortlessly arises whether there is a signal available which can be used to monitor the evolution of a single fault activation process, in analogy to the EEG which is used to monitor the evolution of a single seizure activation process. Such a signal exists. Fracture induced EM fields allow a real-time monitoring of damage evolution in materials during mechanical loading. Crack propagation is the basic mechanism of material failure. EM emissions in a wide frequency spectrum ranging from kHz to MHz are produced by opening cracks, which can be considered as the so-called precursors of general fracture. The radiated EM precursors are detectable both at a laboratory and geological scale \cite{Warwick1982,Gokhberg1982,Hayakawa1994,Hayakawa1999,Hayakawa2002,Nagao2002,Frid2003,Mavromatou2004,Fukui2005,Eftaxias2007a,Uyeda2009,Lacidogna2011,Eftaxias2011}.

As it is said, ``Electromagnetic seismograms'' are fracto-electromagnetic signals which provide us with information about the fracture induced electromagnetic activity \cite{Eftaxias2000,Eftaxias2001a,Eftaxias2002,Eftaxias2003,Eftaxias2004,Kapiris2004,Contoyiannis2005,Karamanos2006,Papadimitriou2008,Kalimeri2008,Contoyiannis2008,Eftaxias2009,Eftaxias2010,Contoyiannis2010}.

The different stages of the earthquake preparation process are reflected in different stages of the emerged EM activity. Indeed, an important feature, observed both at laboratory and geophysical scale, is that the MHz radiation precedes the kHz one \cite{Eftaxias2002}. The remarkable asynchronous appearance of these precursors indicates that they refer to different stages of earthquake preparation process. Moreover, it implies a different mechanism for their origin. The following {\it two stage model of EQ generation by means of pre-fracture EM activities} has been proposed \cite{Kapiris2004,Contoyiannis2005,Karamanos2006,Eftaxias2007a,Papadimitriou2008,Kalimeri2008,Contoyiannis2008,Eftaxias2009,Eftaxias2010,Contoyiannis2010}:

(i) The pre-seismic MHz EM emission is thought to be due to the fracture of the highly heterogeneous system that surrounds the family of large high-strength entities distributed along the fault sustaining the system. The temporal evolution of a MHz EM precursor, which behaves as a second order phase transition, reveals transition from the phase from non-directional almost symmetrical cracking distribution to a directional localized cracking zone that includes the backbone of strong asperities ({\it symmetry breaking}) \cite{Contoyiannis2005}. The identification of the time interval where the {\it symmetry breaking} is completed indicates that the fracture of heterogeneous system in the focal area has been obstructed along the backbone of asperities that sustain the system: {\it The siege of strong asperities begins}. However, the prepared EQ will occur if and when the local stress exceeds fracture stresses of asperities. Consequently, the appearance of a really seismogenic MHz EM anomaly does not mean that the EQ is unavoidable.

(ii)	It has been suggested that the lounge of the kHz EM activity shows the fracture of asperities sustaining the fault. Thus, our analysis based on the study of the recorded kHz EM ''seismograms".

We mainly refer to preseismic kHz EM activities associated with the: Athens (Greece) earthquake ($M = 5.9$) that occurred on September 7, 1999 \cite{Eftaxias2001a,Kapiris2004,Contoyiannis2005,Karamanos2006,Eftaxias2007a,Papadimitriou2008}, and L'Aquila (central Italy) earthquake ($M = 6.3$) that occurred on April 6, 2009 \cite{Eftaxias2009,Eftaxias2010,Contoyiannis2010}. These signals have been recorded with a sampling rate of 1 sample/sec.

\section{Fundamentals of nonextensive statistical mechanics}

Perhaps two of the most vivid and richest examples of the dynamics of a complex system at work are the behaviour of brain / earth crust during the epileptic seizure / earthquake generation. A central property of their generation is the occurrence of large-scale collective behaviour with a very rich structure, resulting from repeated nonlinear interactions among the constituents, namely, firing neurons / opening cracks, of the system. Consequently, the nonextensive statistical mechanics \cite{Tsallis2009} is the appropriate framework in order to investigate the process of launch of the two shocks under study.

The thermodynamical concept of entropy was introduced by Clausius in 1865. A few years later, it was shown by Boltzmann that this quantity can be expressed in terms of the probabilities associated with the microscopic configurations of the system. We refer to this fundamental connection as the Boltzmann-Gibbs (BG) entropy, namely (in its discrete form) $S_{BG} = -k \sum_{i=1}^{W} p_{i} \ln (p_{i})$, where $k$ is the Boltzmann constant, and ${p_i}$ the probabilities corresponding to the $W$ microscopic configurations (hence $\sum_{i=1}^{W} p_{i} = 1$). This entropic form, further discussed by Gibbs, Neumann and Shannon, and constituting the basis of the celebrated BG statistical mechanics, is additive. Indeed, for a system composed of any two (probabilistically) independent subsystems, the entropy $S_{BG}$ of the sum coincides with the sum of entropies. If, on the contrary, the correlations between the subsystems are strong enough, then the additivity of $S_{BG}$ is lost, being therefore incompatible with classical thermodynamics. In such a case, the many and precious relations described in textbooks of thermodynamics become invalid. Along a line which will be shown to overcome this difficulty, and which consistently enables the generalization of BG statistical mechanics, it was proposed by Tsallis in 1988 \cite{Tsallis1988,Tsallis2009} the entropy.

\begin{equation}
S_{q}=k\frac{1}{q-1}\left(1-\sum_{i=1}^{W}p_{i}^{q}\right),
\label{eq:t1}
\end{equation}

\noindent $p_{i}$ are the probabilities associated with the microscopic configurations, $W$ is their total number, and $k$ is Boltzmann's constant. The value $q =1$ corresponds to the standard, extensive, BG statistics.

The value of $q$ is a measure of the nonextensivity of the system reflected in the following additivity rule:

\begin{equation}
S_{q}(A+B)=S_{q}(A)+S_{q}(B)+(1-q)S_{q}(A)S_{q}(B).
\label{eq:t2}
\end{equation}

The index $q$ appears to characterise universality classes of nonadditivity, by phrasing this concept to what is done in the standard theory of critical phenomena. Within each class, one expects to find infinitely many dynamical systems \cite{Tsallis2009}.

We clarify that the parameter $q$ itself is not a measure of the complexity of the system but measures the degree of nonextensivity of the system. It is the time variations of the Tsallis entropy for a given $q$ ($S_{q}$) that quantify the dynamic changes of the complexity of the system. Lower $S_{q}$ values characterise the portions of the signal with lower complexity.

\section{Statistical analysis methods}

One of the objectives of the present work is to examine whether common ``pathological'' features characterise both epileptic seizures and preseismic EM activity. An anomaly in a recorded time series is defined as a deviation from normal (background) behaviour. In order to develop a quantitative identification of an emerged shock, tools of information theory and concepts of entropy rooted in extensive and nonextensive statistical mechanics can be used in order to recognize significant change in the statistical pattern. Entropy and information are seen to be complementary quantities, in a sense: entropy "drops" have as counterpart information "peaks", both indicating a more ordered state / lower complexity. A part of the employed measures (Tsallis entropy, Fisher Information, T-entropy, Block-entropies) are used here in the context of symbolic dynamics. Other measures (approximate entropy, R/S analysis) refer to the raw data. In the following, for the scale of completeness and for later use, a short introduction to symbolic dynamics is provided. It is important to note that one cannot find an optimum organization or complexity measure. Thus, a combination of some such quantities which refer to different aspects, such as structural or dynamical properties, is the most promising way.

\subsection{Fundamentals of symbolic dynamics}

Symbolic time series analysis is a useful tool for modelling and characterisation of nonlinear dynamical
 \cite{Voss1996}. It is a way of coarse-graining or simplifying the description \cite{Hao1989}.

In the framework of symbolic dynamics, time series are transformed into a series of symbols by using an appropriate partition which results in relatively few symbols. After symbolization, the next step is the construction of sequences of symbols (``words'' in the language of symbolic dynamics) from the series of symbols by collecting groups of symbols together in temporal order.

To be more precise, the simplest possible coarse-graining of a time series is given by choosing a threshold $C$ (usually the mean value of the data considered) and assigning the symbols ``1'' and ``0'' to the signal, depending on whether it is above or below the threshold (binary partition). Thus, we generate a symbolic time series from a 2-letter ($\lambda=2$) alphabet (0,1), e.g. $0110100110010110\ldots$. We read this symbolic sequence in terms of distinct consecutive ``blocks'' (words) of length $n=2$. In this case one obtains $01/10/10/01/10/01/01/10/\ldots$. We call this reading procedure ``lumping''. The number of all possible kinds of words is $\lambda^{n}=2^{2}=4$, namely 00, 01, 10, 11. The required probabilities for the estimation of an entropy, $p_{00}$, $p_{01}$, $p_{10}$, $p_{11}$ are the fractions of the blocks (words) 00, 01, 10, 11 in the symbolic time series, namely, 0, 4/16, 4/16, and 0, correspondingly. Based on these probabilities we can estimate, for example, the probabilistic entropy measure $H_S$ introduced by \cite{Shannon1948},

\begin{equation}
H_S = - \sum p_i \ln p_i \,
\end{equation}

where $p_{i}$ are the probabilities associated with the microscopic configurations.

In a symbolic time series of $W$ symbols, $\{A_i\}, i=1,2,...,W$, one can read it by words of length $L=n,\text{ }\left( n<W \right)$. For each word length, there are ${{\lambda }^{n}}$ possible combinations of the symbols that may be found in a word, here ${{\lambda }^{n}}={{2}^{n}}$ , since $\lambda =2$. The probability of occurrence $p_{j}^{(n)}$ of the $j-\text{th}$ combination of symbols ($j={{1,2,...,2}^{n}}$) in a word of length $n$, can be denoted as: 

\begin{equation}
\frac{\text{ }\!\!\#\!\!\text{  of the }j-th\text{ combination found in words of length }n}{\text{total  }\!\!\#\!\!\text{  of words of length }n\text{ (by lumping)}}
\label{eq:probs}
\end{equation}

Based on these probabilities we can estimate, the probabilistic entropy or information measures.

\subsection{Tsallis entropy}

The Tsallis entropy for the word length $n$, $S_{q}(n)$,is
\begin{equation}
{{S}_{q}}\left( n \right)=k\frac{1}{q-1}\left( 1-\sum\limits_{j=1}^{{{2}^{n}}}{{{\left[ p_{j}^{(n)} \right]}^{q}}} \right)
\end{equation}

Broad symbol-sequence frequency distributions produce high entropy values, indicating a low degree of organization. Conversely, when certain sequences exhibits high frequencies, low values are produced, indicating a high degree of organization.

We clarify that the real Tsallis entropy corresponds to the optimal partition. The optimal partition is the one that maximizes the Tsallis-entropy. The corresponding entropy-like quantities for the other partitions are pseudo-Tsallis entropies. For this purpose, the threshold $C$ is initially fixed to the mean value of the data in the particular time window under study. For the corresponding symbolic sequence we estimate the associated ``pseudo-Tsallis entropy''. We repeat the above procedure by changing the threshold $C$ around the mean value. Our analysis indicates that the optimal partition corresponds always to a threshold not very far from the mean value of the segment \cite{Karamanos2006}.

\subsection{Fisher Information}

Fisher Information was first introduced by Fisher \cite{Fisher1925} as a representation of the amount of information that can be extracted from a set of measurements (or the ``quality'' of the measurements) \cite{Mayer2006}. Moreover, it is a powerful tool to investigate complex and non-stationary signals \cite{Martin1999,Telesca2011}. It allows to accurately describe the behaviour of dynamic systems, and to characterise the complex signals generated by these systems \cite{Martin1999,Vignat2003}. It has been used as a measure of the state of disorder of a system or phenomenon, behaving inversely to entropy, i.e., when order increases, entropy decreases, while Fisher Information increases \cite{Mayer2006,Frieden1998}. Furthermore, Fisher Information presents the so called ``locality'' property in contrast to the ``globality'' of entropy (or Shannon's information), referring to the sensitivity of Fisher Information in changes in the shape of the probability distribution corresponding to the measured variable, not presented by entropy \cite{Mayer2006,Frieden2004,Fath2004}.

The Symbolic Fisher Information Measure (SFIM), $I\left( n \right)$, for the word length $n$, is defined in terms of the probability of occurrence $p_{j}^{(n)}$ of the $j-\text{th}$ combination of symbols ($j={{1,2,...,2}^{n}}$) in a word of length $n$ of Eq. (\ref{eq:probs}), as \cite{Potirakis2011a}: 

\begin{equation}
I\left( n \right)=\sum\limits_{j=1}^{{{2}^{n}}-1}{\frac{{{\left[ p_{j+1}^{(n)}-p_{j}^{(n)} \right]}^{2}}}{p_{j}^{(n)}}}
\end{equation}

\subsection{T-entropy of a string}

$T$-entropy is a novel grammar-based complexity / information measure defined for finite strings of symbols \cite{Ebeling2001,Titchener2005}. It is a weighted count of the number of production steps required to construct a string from its alphabet. {\it Briefly, it is based on the intellectual economy one makes when rewriting a string according to some rules}. An example of an actual calculation of the $T$-entropy for a finite string is given in \cite{Ebeling2001,Karamanos2006}.

\subsection{The Shannon-like n-block entropies}

Block entropies, depending on the word-frequency distribution, are of special interest, extending Shannon's classical definition of the entropy of a single state \cite{Shannon1948} to the entropy of a succession of states \cite{Nicolis1994}. Shannon $n$-block entropies (conditional entropy, entropy of the source, Kolmogorov-Sinai entropy) measure the uncertainty of predicting a state in the future, provided a history of the present state and the previous states \cite{Eftaxias2009}.

(i) {\it The Shannon $n$-block entropies}

Following Shannon's approach \cite{Shannon1948} the $n$-block entropy, $H(n)$, is given by
\begin{equation}
H\left( n \right)=-\sum\limits_{j=1}^{{{2}^{n}}}{p_{j}^{(n)}\ln }p_{j}^{(n)}
\end{equation}
The $H(n)$ is a measure of uncertainty and gives the average amount of information necessary to predict a sub-sequence of length $n$.

(ii) {\it The Shannon $n$-block entropy per letter}

This entropy is defined by
\begin{equation}
h^{n} = \frac{H(n)}{n}.
\end{equation}
This entropy may be interpreted as the average uncertainty per letter of an $n$-block.

(iii) {\it The conditional entropy}

From the Shannon $n$-block entropies we derive the conditional (dynamic) entropies by the definition

\begin{equation}
h_{n} = H(n+1) - H(n).
\end{equation}
The conditional entropy $h_n$ measures the uncertainty of predicting a state one step into the future, provided a history exists of the preceding n states.

Predictability is measured by conditional entropies. For Bernoulli sequences we have the maximal uncertainty

\begin{equation}
h_{n} = log(\lambda).
\end{equation}

Therefore we define the difference

\begin{equation}
r_n=log(\lambda)-h_n
\end{equation}

as the average predictability of the state following a measured $n$-trajectory. In other words, predictability is the information we get by exploration of the next state in comparison to the available knowledge. We use, in most cases, $\lambda$ as the base of the logarithm. Using this base, the maximal uncertainty/predictability is one \cite{Eftaxias2009}. In general our expectation is that any long-range memory decreases the conditional entropies and improves our chances for prediction.

(iv) {\it The entropy of the source}

A quantity of particular interest is the entropy of the source, defined as

\begin{equation}
h = \lim_{n \rightarrow \infty} h_{(n)} = \lim_{n \rightarrow \infty} h^{(n)}
\end{equation}

The limit entropy $h$ is the discrete analog of Kolmogorov-Sinai entropy. {\it It is the average amount of information necessary to predict the next symbol when being informed about the complete pre-history of the system}. Since positive Kolmogorov-Sinai entropy implies the existence of a positive Lyapunov exponent, it is an important measure of chaos.

\subsection{Approximate entropy}

Related to time series analysis, the approximate entropy $ApEn$ provides a measure of the degree of irregularity or randomness within a series of data (of length $N$). $ApEn$ was pioneered by Pincus as a measure of system complexity \cite{Pincus1991}. It was introduced as a quantification of regularity in relatively short and noisy data. It is rooted in the work of Grassberger and Procaccia \cite{Grassberger1983} and has been widely applied to biological systems \cite{Pincus1994,Pincus1996}. The approximate entropy examines time series for similar epochs: more similar and more frequent epochs lead to lower values of $ApEn$. In summary, $ApEn$ is a ``regularity statistics'' that quantifies the unpredictability of fluctuations in a time series. The presence of repetitive patterns of fluctuation in a time series renders it more predictable than a time series in which such patterns are absent. A time series containing many repetitive patterns has a relatively small $ApEn$; a less predictable (i.e., more complex) process has a higher $ApEn$. An example of an actual calculation of the approximate entropy is given in \cite{Karamanos2006}.

\subsection{R/S analysis}

The Rescaled Range Analysis ($R/S$), which was introduced by Hurst \cite{Hurst1951}, attempts to find patterns that might repeat in the future. Briefly, there are two main variables used in this method, namely, the range of the data, $R$, as it is measured by the highest and lowest values in the time period, and the standard deviation of the data $S$. $R/S$ is expected to show a power-law dependence on the bin size $n$:
\begin{equation}
R(n)/S(n) \sim n^H ,
\end{equation}

where $H$ is the Hurst exponent. The range $0.5 < H < 1$ (indicates persistency, which means that if the amplitude of the fluctuations increases in a time interval it is likely to continue increasing in the next interval. The range $0 < H < 0.5$ indicates antipersistency, which means that if the amplitude of the fluctuations increases in a time interval it is likely to continue decreasing in the next interval. An example of an actual calculation of the approximate entropy is given in \cite{Eftaxias2009}.

\section{Application of statistical analysis methods to data}

In this section, we examine the epileptic and preseismic time series in terms of the above mentioned multidisciplinary statistical procedure, aiming to discover possible common ``pathological'' symptoms of transition from the normal (``healthy'' stage) to a significant seismic or epileptic rat or human shock (``pathological'' stage). We show that similar distinctive symptoms accompany the appearance of the biological and geophysical crises under study, which sensitively recognize and discriminate each of them from the corresponding background ``noise'': the transition from antipersistent to persistent behaviour and sift to a significantly higher organization indicates the onset of two crises. The appearance of a high organization dynamics which is simultaneously characterised by a positive feedback mechanism is consistent with the emergence of a catastrophic phenomenon.

A way to examine transient phenomena in a time-series is to analyze it into a sequence of distinct time windows of short duration and compute the various measures of the degree of organisation/information content in each one of them.

\subsection{Comparison between epileptic seizures of rats and preseismic electromagnetic emissions}

\begin{figure}[h]
\begin{center}
\includegraphics[width=0.5\textwidth]{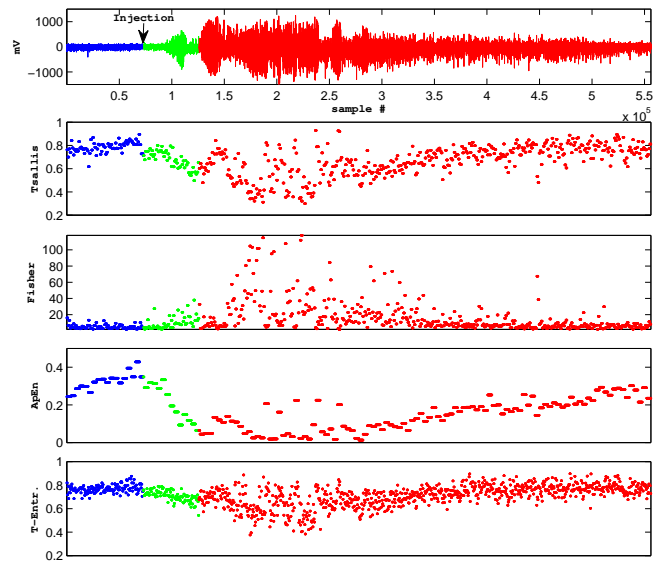}
\end{center}
\caption{The upper part depicts a rat EEG containing an evoked seizure. The arrow shows the time of injection. The blue, green and red parts refer to the normal, pre-ictal and seizure epochs, correspondingly. The next sub-figures show the temporal evolution of Tsallis entropy, Fisher information, Approximate entropy and T-entropy, respectively, using fixed windows of 1024 samples each.}
\label{fig:EEG_entropies}
\end{figure}

Fig. \ref{fig:EEG_entropies} shows an EEG signal of a rat, recorded during an evoked seizure, along with its analysis in terms of Tsallis entropy, Approximate entropy, T-entropy, and Fisher Information. Three distinct phases of signal have been identified by all the employed metrics. The first (blue) part is the normal state, before the applied injection. It is followed by the pre-ictal (green) phase and then by the actual seizure (red) phase.

Tsallis entropy in its symbolic form for word length 5 and $q=1.7$ (see Sec. 4) shows that the underlying process becomes progressively more organized as one moves from the normal state to the seizure state. Approximate entropy further verifies the previous result by detecting more similar and more frequent epochs as one moves to the seizure part of the signal, while T-entropy also agree detecting the lowest number of required steps to construct the corresponding symbolic array during the seizure phase of the signal. Accordingly, Fisher Information, shows that higher information content, than that of the normal phase, can be detected in both the subsequent phases, but the maximum information content is within the seizure phase.

\begin{figure}[h]
\begin{center}
\includegraphics[width=0.5\textwidth]{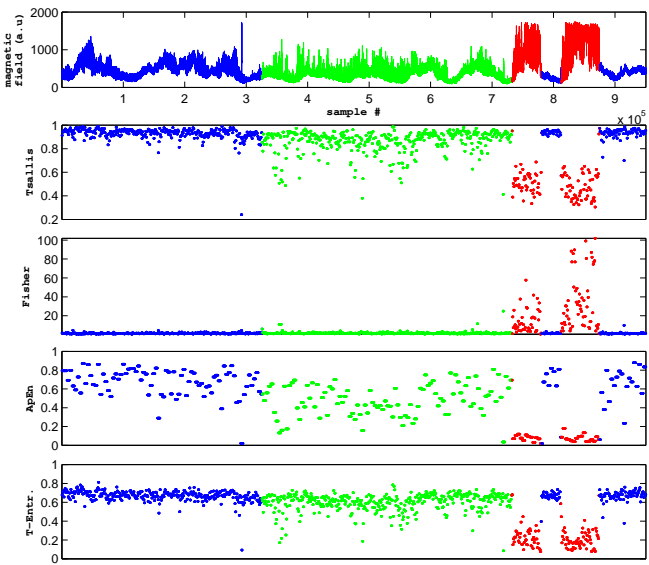}
\end{center}
\caption{The upper part depicts the EM time series associated with the Athens earthquake recorded by the 10 kHz magnetic sensor. The blue part refers to the background noise. The green and red parts refer to the two distinct epochs of the emerged preseismic EM activity (see text). The next sub-figures show the temporal evolution of Tsallis entropy, Fisher information, Approximate entropy and T-entropy, respectively, using fixed windows of 1024 samples each.}
\label{fig:EM_entropies}
\end{figure}

Fig. \ref{fig:EM_entropies} shows the same analyses as those reported for the rat seizure (Fig. \ref{fig:EM_entropies}) for the 10 kHz preseismic EM time series associated with the Athens earthquake. Three distinct parts can be identified by all the employed metrics, as it is happened in the case of the rat EEG. The first (blue) part is the background noise. The second (green) and third (red) parts refer to two distinct stages of the preseismic EM activity \cite{Eftaxias2003,Eftaxias2004,Kapiris2004,Contoyiannis2005,Karamanos2006,Papadimitriou2008}.

All the entropic and information measures show that the underlying process becomes progressively more organized. The green epoch of the EM precursor behaves as the pre-ictal phase of the rat EEG (Fig. \ref{fig:EEG_entropies}), namely, it is characterised by a population of EM events sparsely distributed in time with noteworthy higher order of organization content in comparison to that of the noise. The abruptly emerged two strong EM bursts A and B, which are included in red epoch, are characterised by a significantly higher organization and information content in comparison to those of the green epoch. We observe a strong analogy in terms of order of organization between the two abruptly emerged strong EM bursts and the abruptly emerged rat seizure.

\begin{figure}[h] 
\begin{center}
\includegraphics[width=0.5\textwidth]{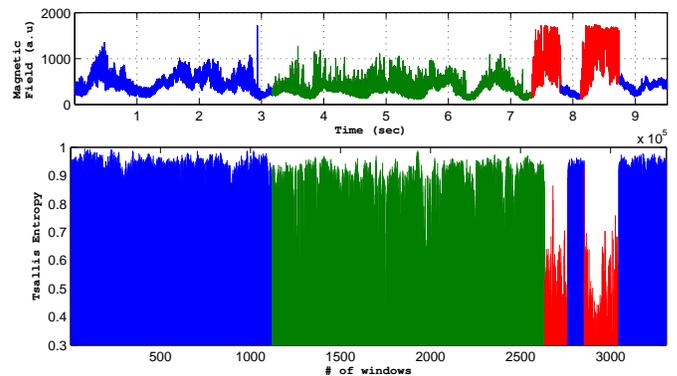}
\end{center}
\caption{EM time series recorded by the 10kHz sensor associated with the Athens EQ (upper graph). Tsallis entropy (bottom graph), for the population of the non-overlapping data windows of 1024 samples each, which are proved to be stationary. The nonextensive parameter $q=1.8$ was used (see sec. 7).}
\label{fig:EM_entropy_stat}
\end{figure}

{\it Remark I}: It should be noted here that the Tsallis entropy calculations presented in Figs \ref{fig:EEG_entropies} and \ref{fig:EM_entropies} were performed using non-overlapping windows of 1024 points without check of stationarity, considering that the window length is small enough to provide preudo-stationarity conditions. Fig. \ref{fig:EM_entropy_stat} presents the Tsallis entropy calculation solely on the population of the windows which were proven to be stationary. The results shown in Fig. \ref{fig:EM_entropy_stat} are the same to those presented in Fig. \ref{fig:EM_entropies}.

{\it Remark II}: We refer to the above mentioned proposal that the two strong EM bursts which present the highest order of organization and maximum information content, correspond to the epileptic rat seizure. The first EM burst contains approximately 20\% of the total EM energy received and the second one the remaining 80\% \cite{Eftaxias2001a}. On the other hand, the fault modelling of the Athens EQ, based on information obtained by radar interferometry, predicts two faults: the main fault segment is responsible for 80\% of the total energy released, while the secondary fault segment for the remaining 20\% \cite{Eftaxias2001a}. The last surprising experimental correlation supports the hypothesis that the two strong kHz EM bursts were sourced in the nucleation of the impending earthquake \cite{Eftaxias2001a,Kapiris2004,Contoyiannis2005,Karamanos2006,Papadimitriou2008}. This experimental evidence justifies our proposal.

\begin{figure}[h]
\begin{center}
\includegraphics[width=0.5\textwidth]{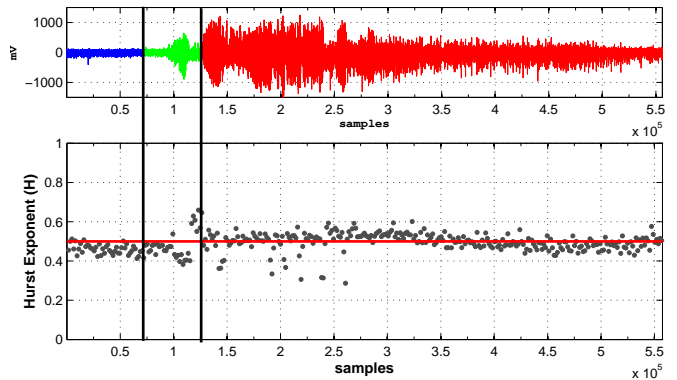}
\end{center}
\caption{Rat EEG containing an evoked seizure (upper graph), showing the time of injection. The lower part depicts the temporal evolution of the Hurst exponent obtained through R/S analysis, using fixed windows of 1600 samples each.}
\label{fig:RAT_Hurst}
\end{figure}

{\it Remark III}: We underline that the green and red epochs of the EM precursor (see Fig. \ref{fig:EM_entropies}) are characterised by antipersistency and persistency, correspondingly \cite{Eftaxias2003,Eftaxias2004,Kapiris2004,Contoyiannis2005,Karamanos2006}. In direct analogy, the rat EEG also follows the aforementioned crucial scheme: the pre-ictal phase shows antipersistent behaviour ($H<0.5$), while the phase of seizure is characterised by persistency ($H>0.5$) (Fig. \ref{fig:RAT_Hurst}). This scheme has been also verified in terms of fractal spectral analysis \cite{Li2005,Kapiris2005,Eftaxias2006}. The above mentioned results enhance the proposal that the green pre-ictal phase of rat EEG (Fig. \ref{fig:EEG_entropies}) corresponds to the initial green epoch of the emerged EM precursor.

\subsection{Comparison between human epileptic seizures and preseismic electromagnetic emissions}

In the prospect to examine the repeatability of the results that concern this field of research, we focus on the comparison between human EEGs \cite{Andrzejak2001} and EM precursors associated with large earthquakes.

\begin{figure}[h]
\begin{center}
\subfloat []{\label{subfig:humanEEGens}\includegraphics[width=0.5\textwidth]{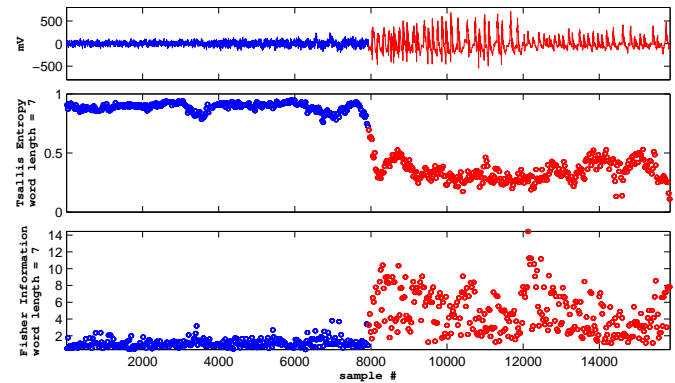}}\\
\subfloat []{\label{subfig:fourEEGs}\includegraphics[width=0.5\textwidth]{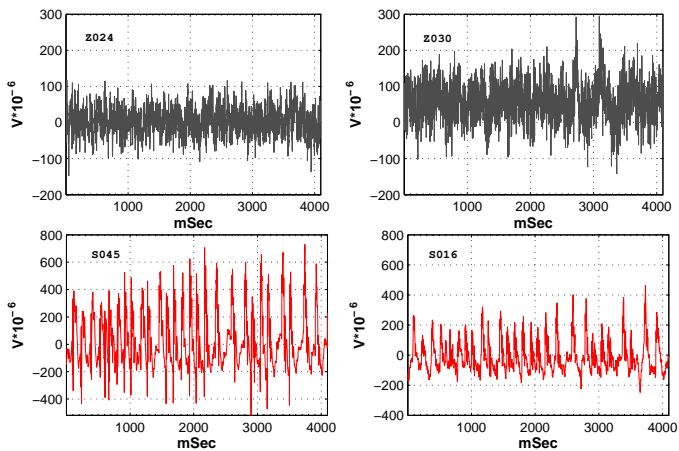}}
\end{center}
\caption{(a) The upper part depicts a sequence of two healthy human EEGs and a sequence of two human epileptic seizures. The healthy EEGs are blue coloured whereas the patient EEGs are red coloured. In the lower parts the temporal evolution of Tsallis entropy and Fisher information is presented. (b) depicts a zoomed version of the four EEGs used.}
\end{figure}

In Fig. \ref{subfig:humanEEGens} we present four human EEG signals analysed in terms of Tsallis entropy and Fisher information using fixed windows of 1024 samples each. The blue graphs refer to a sequence of two healthy EEGs included in the set ``A'', while the red ones to a sequence of two epileptic seizures included in the set ``B'' \cite{Andrzejak2001}. The results of this analysis reveal that the patient EEGs are distinguished by a higher order of organization and higher information content than the healthy ones.

\begin{figure}[h]
\begin{center}
\includegraphics[width=0.5\textwidth]{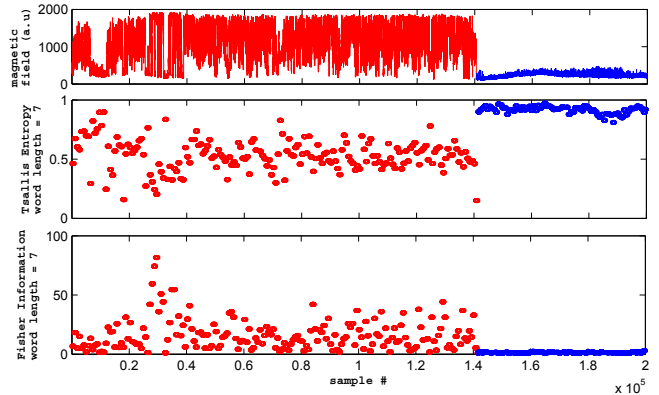}
\end{center}
\caption{In the upper part the emerged EM precursor (red time series) in the case of the Methoni earthquake is presented. This precursor ceased about 4 days before the earthquake occurrence. The blue time series refers to the background noise. In the lower parts depict the temporal evolution of Tsallis entropy and Fisher information.}
\label{fig:methoni_ens}
\end{figure}

In Fig. \ref{fig:methoni_ens} we present the above analysis applied to the kHz EM precursor associated with the Methoni (Greece) earthquake which occurred on February 14, 2008 with magnitude $M = 6.7$. Herein we mention that the EM precursor (red part) was stopped approximately 4 days before the earthquake occurrence. The blue part refers to the background noise. The analysis in terms of Tsallis entropy and Fisher information shows that the EM precursor (red part) is distinguished by a higher order of organization along with a higher information content in respect to the background noise. We note that the two human epileptic seizures and the EM precursor, follow a persistent behaviour.

\begin{figure}[h]
\begin{center}
\includegraphics[width=0.5\textwidth]{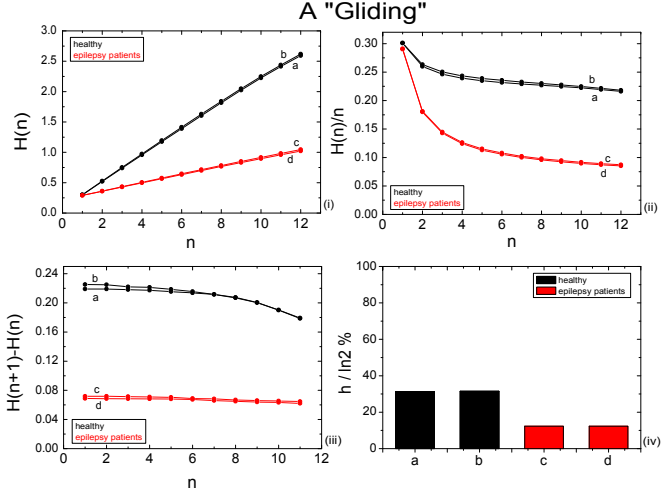}
\end{center}
\caption{Block entropies for the four human EEGs of Fig. \ref{subfig:fourEEGs}. The healthy EEGs which are black colored are characterised by higher complexity/lower predictability in respect to that of the epileptic seizures (red coloured).}
\label{fig:humanEEGs_bens}
\end{figure}

\begin{figure}[h]
\begin{center}
\includegraphics[width=0.5\textwidth]{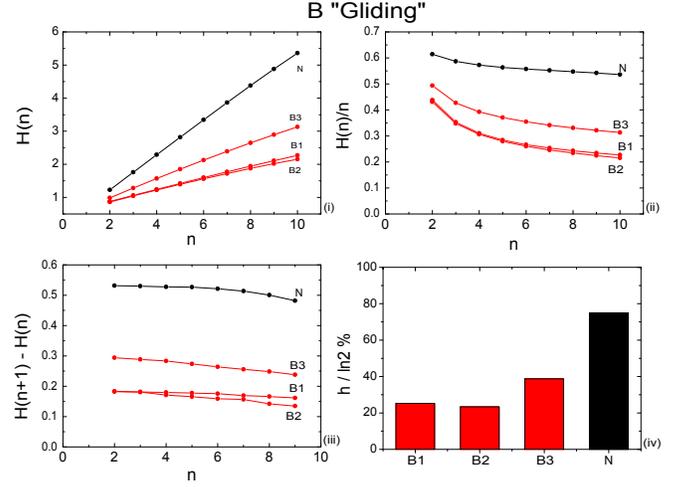}
\end{center}
\caption{Block entropies for the four selective windows of EM time series associated with the L'Aquila earthquake. The black curves refer to the EM background noise while the red curves refer to the three distinct EM bursts emerged prior to L'Aquila earthquake. We observe that these three EM precursors are characterised by lower complexity/higher predictability in respect to that of the back ground noise.}
\label{fig:LaquilaEM_bens}
\end{figure}

The same correspondence between seizure and fault activation is verified by the comparison between the four human EEG signals depicted in Fig. \ref{subfig:fourEEGs} and the pre-seismic EM emissions recorded prior to the L'Aquila EQ \cite{Eftaxias2009,Eftaxias2010,Contoyiannis2010} in terms of block entropies. Figs. \ref{fig:humanEEGs_bens} and \ref{fig:LaquilaEM_bens} show the analyses in terms of block entropies for the four human EEGs and four selective windows of EM time series associated with the L'Aquila earthquake, respectively. All metrics, show that the ``healthy'' (black) windows, present higher entropy values and thus higher degree of randomness than the ``patient'' (red) windows which are clearly more organized. This is valid both for the EEG signals and the EM signals.

\begin{figure}[h]
\begin{center}
\subfloat {\includegraphics[width=0.5\textwidth]{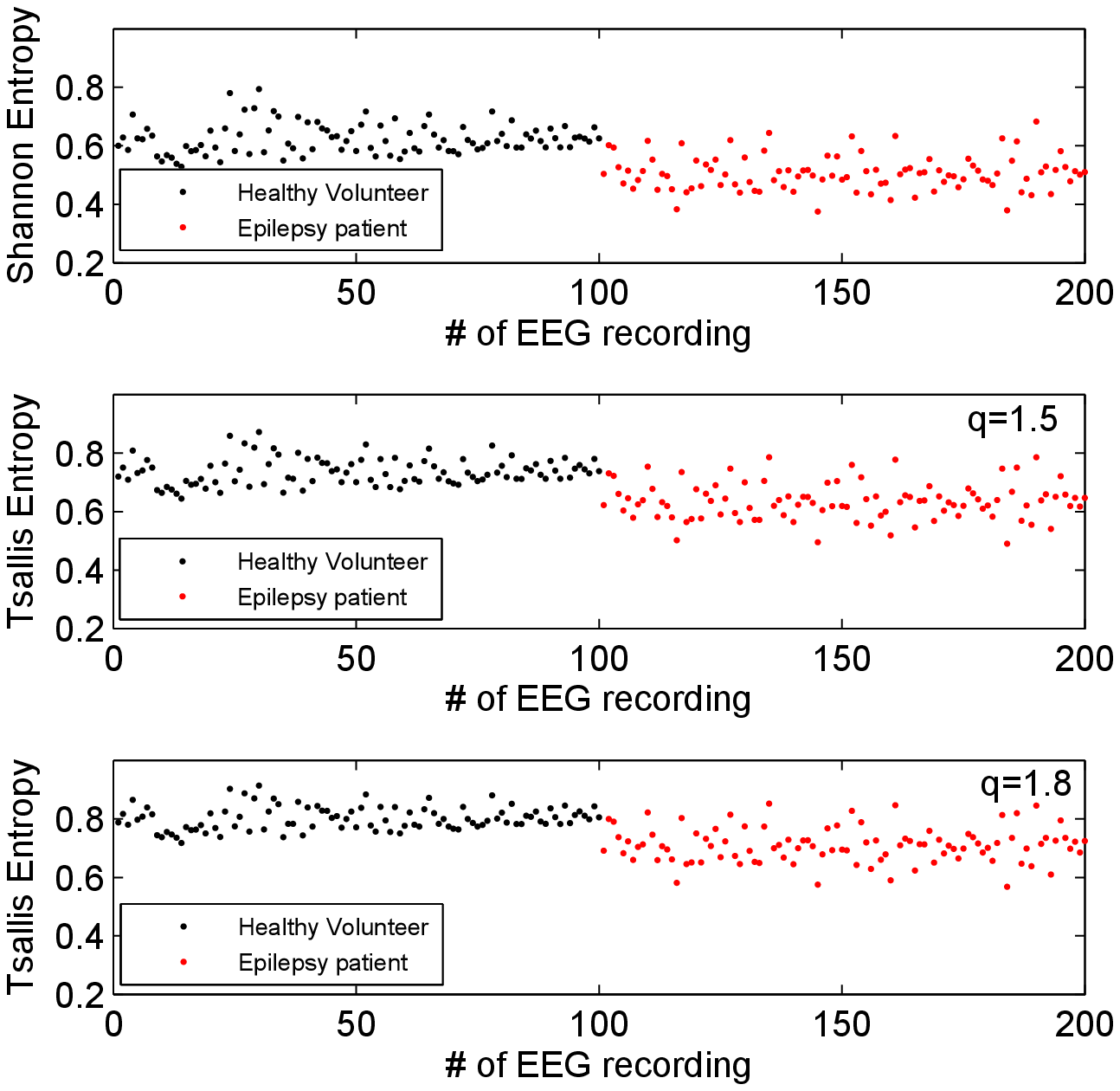}}

\subfloat {\includegraphics[width=0.5\textwidth]{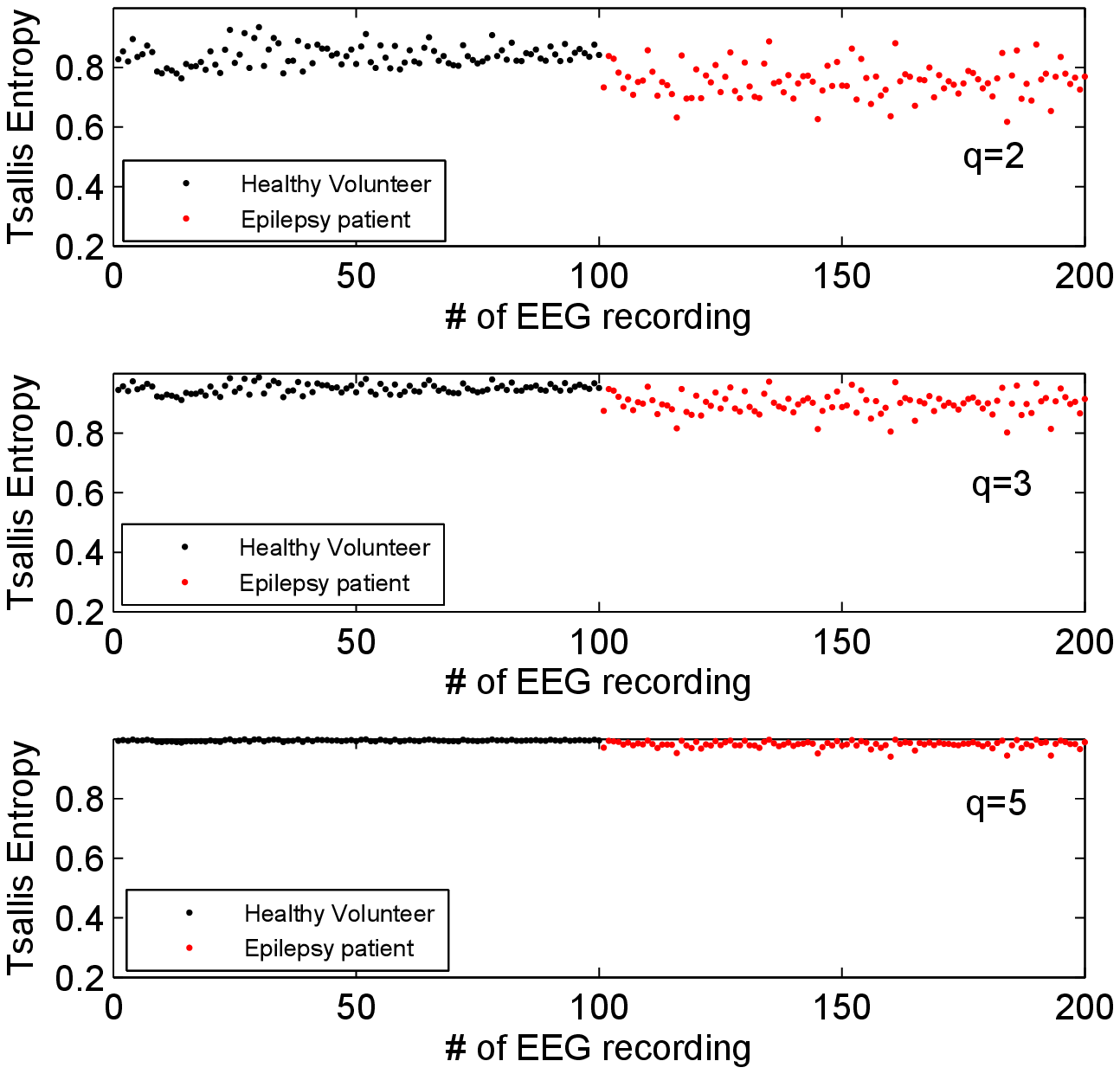}}
\end{center}
\caption{Shannon entropy (top-left), and Tsallis entropy of 100 healthy and 100 patient EEGs belonging to the ``A'' and ``B'' set, correspondingly \cite{Andrzejak2001}, for different values of $q$. The healthy EEGs are black colored while the patient EEGs are red coloured.}
\label{fig:100EEGs_ens}
\end{figure}

\begin{figure}[h]
\begin{center}
\subfloat []{\includegraphics[width=0.5\textwidth]{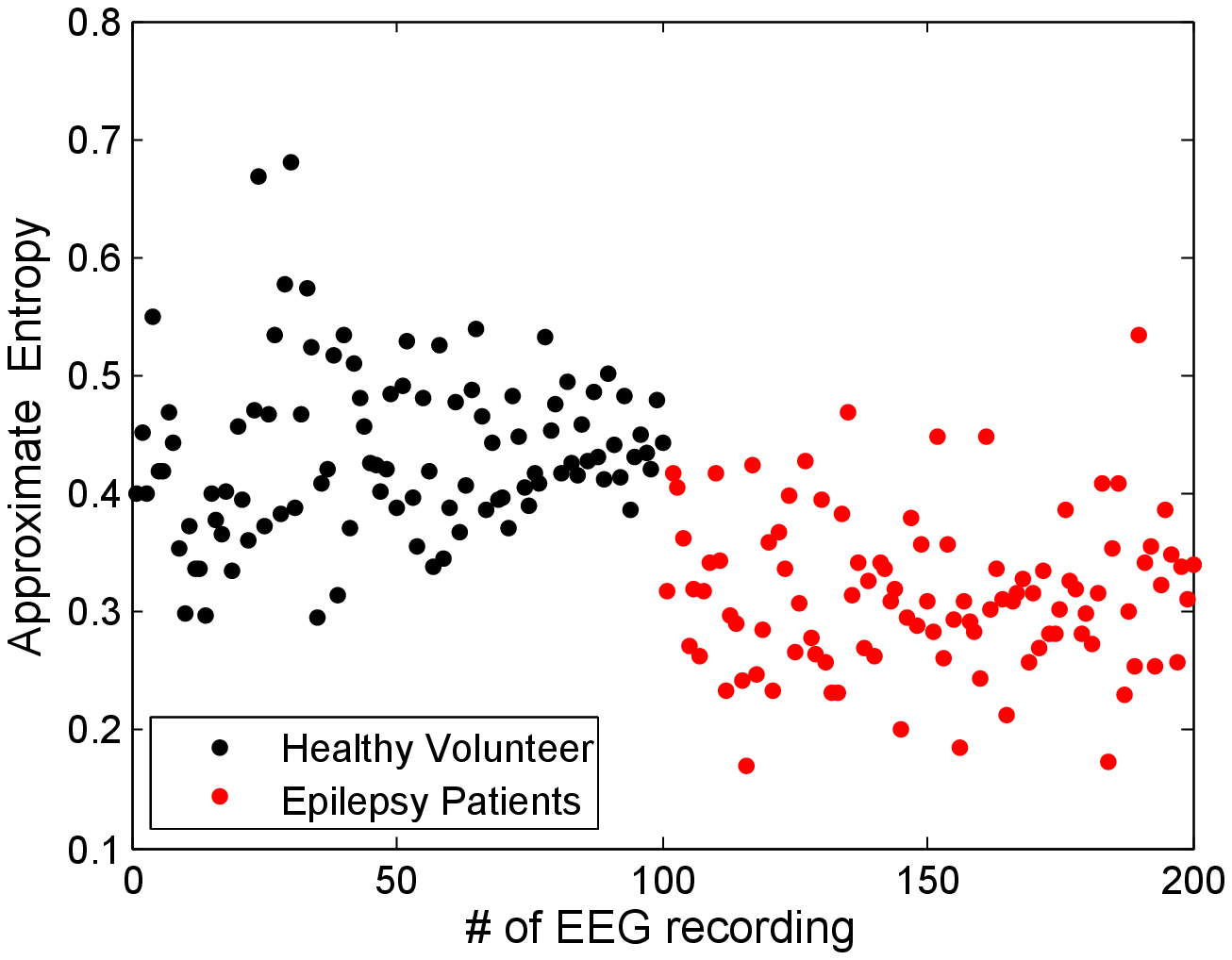}}

\subfloat []{\includegraphics[width=0.5\textwidth]{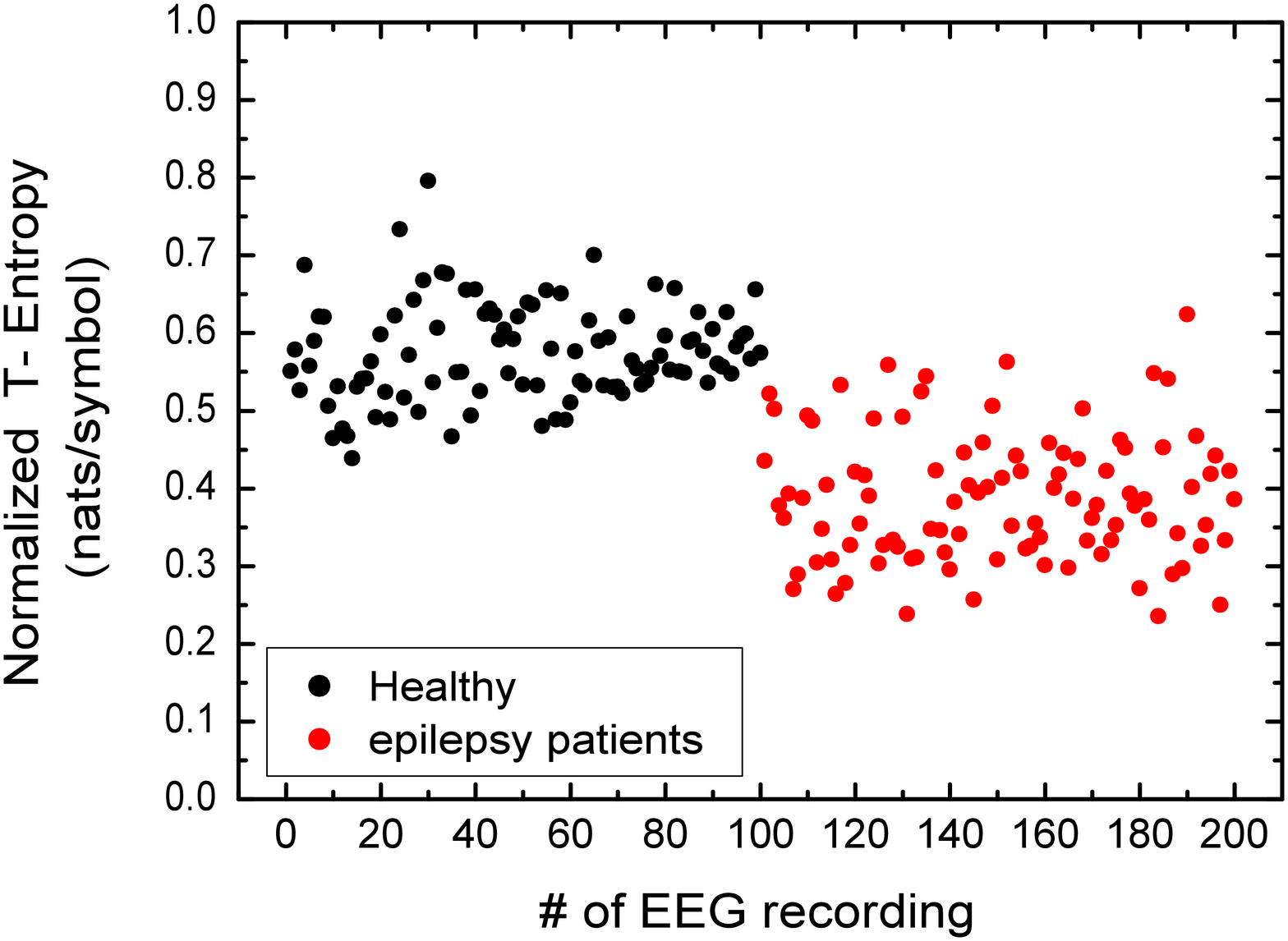}}
\end{center}
\caption{(a) Approximate entropy and (b) T-entropy of 100 healthy (belonging to the ``A'' set) and 100 patient (belonging to the ``E'' set) EEGs \cite{Andrzejak2001}, respectively. The healthy EEGs are black colored while the patient EEGs are red coloured.}
\label{fig:100EEGs_apen_ten}
\end{figure}

It should be stressed out that the validity of these results is not limited to the two individual cases of human seizures presented above; the observed significant increase of the order of organization is a coherent finding among numerous cases. Indeed, the higher degree of organization of the patient EEGs in relation to the healthy ones, is verified by analyzing 100 healthy and 100 human patient EEGs included in the sets ``A'' and ``E'' \cite{Andrzejak2001}. Methods used for this analysis include: Tsallis Entropy (Fig. \ref{fig:100EEGs_ens}), Approximate and T-entropy (Fig. \ref{fig:100EEGs_apen_ten}).

{\it Remark}: As expected, our results depend upon the Tsallis q-value. Fig. \ref{fig:100EEGs_ens} clearly illustrates the superiority of the $q$-values restricted in the range $1 < q < 2$ to magnify differences of the order of organization between healthy and patient EEGs. It is worth mentioning that the nonadditive $q$-parameters that clearly quantify the degree of organization in the EEG time series, are in full agreement with the upper limit $q < 2$ obtained from several studies involving the Tsallis nonadditive framework \cite{Kalimeri2008}. Moreover, this ranging of the $q$- parameter is in harmony with an underlying sub-extensive system, $q > 1$, verifying multiple interactions and information transitions at work during the emergence of the seizure. In Section 7 we show that the system under study is characterised by $q \sim 1.6$.

In the following sections we examine the existence of dynamical correspondences between seizure and earthquake generation by means of scale-free statistics, namely, a common hierarchical organization that results in power-law behaviour over a wide range of values of some parameter such as event energy or waiting time. We concentrate on the question of whether the corresponding power-laws, if any, share the same exponent.

\section{Dynamical analogy by means of ``Gutenberg-Richter law''}

The best known scaling relation for earthquakes is the Gutenberg-Richter (G-R) magnitude-frequency relationship

\begin{equation}
\log N(>m)=\alpha-bm,
\end{equation}

where $N(m>)$ is the cumulative number of earthquakes with a magnitude greater than $m$ occurring in a specified time and area included many faults.
The parameters $b$ and $\alpha$, are constants. The constant $\alpha$ is a measure of the regional level of seismicity. This relation is valid for earthquakes both regionally and globally. In parallel, the probability density function for having an earthquake energy $E$ is denoted by the power-law $P(E)\sim E^{-B}$ where $B \sim 1.4-1.6$.

The probability of an event (seizure or earthquake) having energy $E$ is proportional to $x^{-B}$, where $B \sim 1.5-1.7$ \cite{Osorio2010}. We examine whether the sequences of electrical pulses / EM pulses included in a single seizure / single EM precursor also follow a power-law $P(E)\sim E^{-B}$, with a similar exponent.

\begin{figure}[h]
\begin{center}
\includegraphics[width=0.45\textwidth]{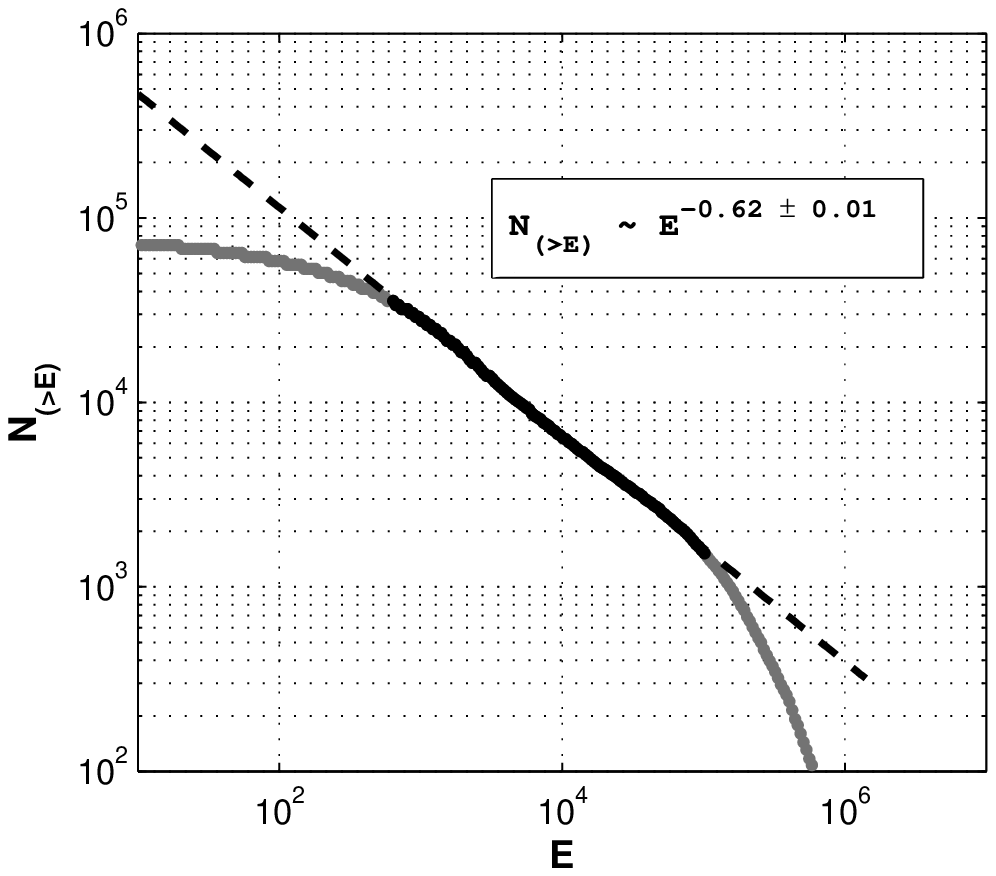}
\end{center}
\caption{The distribution of energies, $E$, of the electrical pulses included in the rat seizure depicted in Fig. \ref{fig:EEG_entropies} follows the power-law $N(>E) \sim E^{-0.62}$.}
\label{fig:endist_rat}
\end{figure}

Fig. \ref{fig:endist_rat} shows that the energies, $E$, of the electrical pulses included in the single rat seizure depicted in Fig. \ref{fig:EEG_entropies} follows the power-law $N(>E) \sim E^{-0.62}$, or equivalently, the power-law $N(E) \sim E^{-1.62}$.

We refer to the case of the kHz EM precursor associated with the Athens earthquake (Fig. \ref{fig:EM_entropies}, upper panel). It has been shown that the cumulative number $N(>A)$ of pre-seismic EM pulses having amplitudes larger than $A$ follows the power-law $N(>A)\sim A^{-0.62}$ \cite{Kapiris2004b}. The probability of an EM-pulse having energy $E$ is proportional to $E^{-1.31}$ \cite{Eftaxias2004}.

The above mentioned results indicate the following peculiarity of dynamical correspondence between seizures and earthquakes. The sequences of: (i) fracto-EM-pulses included in single EM-precursor associated with the activation of a single fault, (ii) electric pulses included in a single seizure, (ii) different earthquakes occurred in areas included many faults, and (iv) different seizures, follow the power-law $P(E)\sim E^{-B}$ with a rather similar $B$-exponent. Notice, in general, differences in constituting elements (organic vs inorganic), in scale, and in other properties between the earth and brain may account for dissimilarities in the values of exponents \cite{Osorio2010}.

The reported dynamical analogy in \cite{Osorio2010} between seizures and earthquakes by means of energy is extended up to the scale of laboratory seismicity. Acoustic / EM emission in rocks has been studied as a model of natural seismicity. Rabinovitch et al. \cite{Rabinovitch2001} have recently studied the fractal nature of EM radiation induced by rock fracture. The analysis of the pre-fracture EM time series reveals that the cumulative distribution function of the amplitudes follows the power $N(>A)\sim A^{-0.62}$, namely, the distribution function of the amplitudes follows the power-law $P(E)\sim E^{-1.31}$. Petri et al. \cite{Petri1994} have performed the statistical analysis of acoustic emission time series in the ultrasonic frequency range, obtained experimentally from laboratory samples subjected to external uni-axial elastic stress. They found a power-law scaling behaviour in the acoustic emission energy distribution with $B=1.3 \pm 0.1$. Houle and Sethna \cite{Houle1996} found that the crumpling of paper generates acoustic pulses with a power-law distribution in energy with $B=1.3-1.6$.

In the next section, we further examine dynamical analogies between seizures and earthquakes in terms of a nonextensive model for earthquake dynamics.

\section {A nonextensive ``Gutenberg-Richter law''}

A model for earthquake dynamics consisting of two rough profiles interacting via fragments filling the gap has been recently introduced by Sotolongo-Costa and Posadas \cite{Sotolongo2004}. The motion of the fault planes can be hindered not only by the overlapping of two irregularities of the profiles, but also by the eventual relative position of several fragments. Thus, the mechanism of triggering earthquakes is established through the combination of the irregularities of the fault planes, on one hand, and the fragments between them, on the other hand. The fragments size distribution function comes from a nonextensive Tsallis formulation, starting from first principles, i.e., a nonextensive formulation of the maximum entropy principle. This nonextensive approach leads to a G-R type law for the magnitude distribution of EQs. Silva et al. \cite{Silva2006} have subsequent revised this model considering the current definition of the mean value, i.e., the so-called $q$-expectation value. The revised model

\begin{eqnarray}
\lefteqn{\log \left[N\left(>M\right)\right]  =} \nonumber\\
&& \log N+\left({2-q\over 1-q} \right)\log \left[1-\left({1-q\over 2-q} \right)\left({10^{2M} \over a^{2/3} } \right)\right] \label{eq:xdef}
\label{eq:silva}
\end{eqnarray}

also provides an excellent fit to seismicities. In Eq. (\ref{eq:silva}) $N$ is the total number of earthquakes, $N_{>M}$ the number of earthquakes with magnitude larger than $M$, and $M\approx \log \varepsilon$. $\alpha$ is the constant of proportionality between the earthquake energy, $\varepsilon$, and the size of fragment. Notice, the $q$-values are distributed within the range of ($1.6 - 1.8$) \cite{Sotolongo2004,Silva2006,Matcharashvili2011,Telesca2010} in various seismic regions under study.

The nonextensive formula (\ref{eq:silva}) also describes the detected EM precursors associated with the activation of a single fault \cite{Papadimitriou2008,Eftaxias2001a}. This finding further supports the self-affine nature of fracture and faulting. We briefly focus on this point.

{\it The notion of ``EM-earthquakes''}: The background (noise) level of the EM time series $A(t_{i})$ is $A_{noise}=500 mV$. We regard as amplitude $A$ of a candidate ``fracto-electromagnetic emission'' the difference $A_{fem}(t_{i})=A(t_{i})-A_{noise}$. We consider that a sequence of $k$ successively emerged ``fracto-electromagnetic emissions'' $A_{fem}(t_{i})$, $i=1,\ldots,k$ represents the EM energy released, $\varepsilon$, during the damage of a fragment. We shall refer to this as an ``electromagnetic earthquake'' (EM-EQ). Since the squared amplitude of the fracto-electromagnetic emissions is proportional to their energy, the magnitude $M$ of the candidate EM-EQ is given by the relation

\begin{equation}
M=\log\varepsilon \sim \log\left(\sum\left [ A_{fem}(t_{i})\right]^{2}\right),
\end{equation}

It has been shown that Eq. (\ref{eq:silva}) provides an excellent fit to the sequence of pre-seismic ``EM-EQs'' associated with the activation of a single fault \cite{Papadimitriou2008,Kalimeri2008,Eftaxias2010}, incorporating the characteristics of nonextensivity statistics into the distribution of the detected precursory EM-EQs. Herein, $N(M>)$ the number of ``EM-EQs'' with magnitude larger than $M$, and $\alpha$ the constant of proportionality between the EM energy released and the size of fragment. The best-fit parameters for this analysis are given by $q \sim 1.80$.

Importantly, it has been shown that the Tsallis-based energy distribution function (Eq. \ref{eq:silva}) is also able to describe solar flares and magnetic storms, as well. The best-fit for this analysis is given by a $q$-parameter value equal $1.82$ and $1.84$, correspondingly \cite{Balasis2011}. It is very interesting to observe the similarity in the $q$-values for: (i) seismicities generated in various large geographic areas, (ii) the precursory sequence of ``EM-EQs'' associated with the activation of a single fault, (iii) solar flares, and (iv) magnetic storms. A common characteristic in the dynamics of the above-mentioned four phenomena is that the energy release is basically fragmentary, i.e., the events being composed of elementary building blocks. The energy release is basically fragmentary in the case of epileptic seizure, as well. Therefore, we examine whether the formula (\ref{eq:silva}) also describes the distribution of energy of the electric fluctuations included in a single seizure.

\begin{figure*}[ht]
\begin{center}
\subfloat []{\label{subfig:silva_100EEGs}\includegraphics[width=0.45\textwidth]{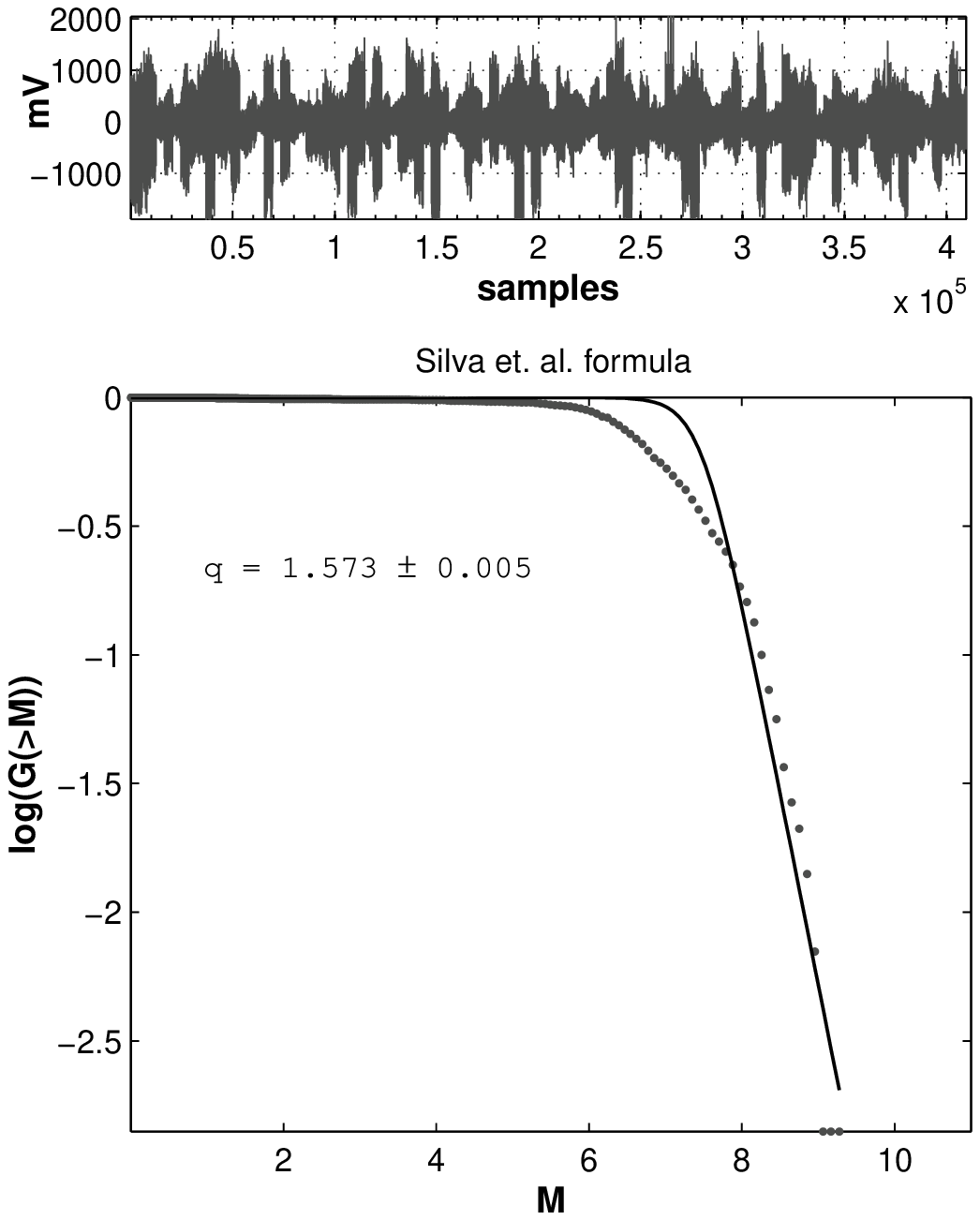}}
\subfloat []{\label{subfig:silva_1EEG}\includegraphics[width=0.45\textwidth]{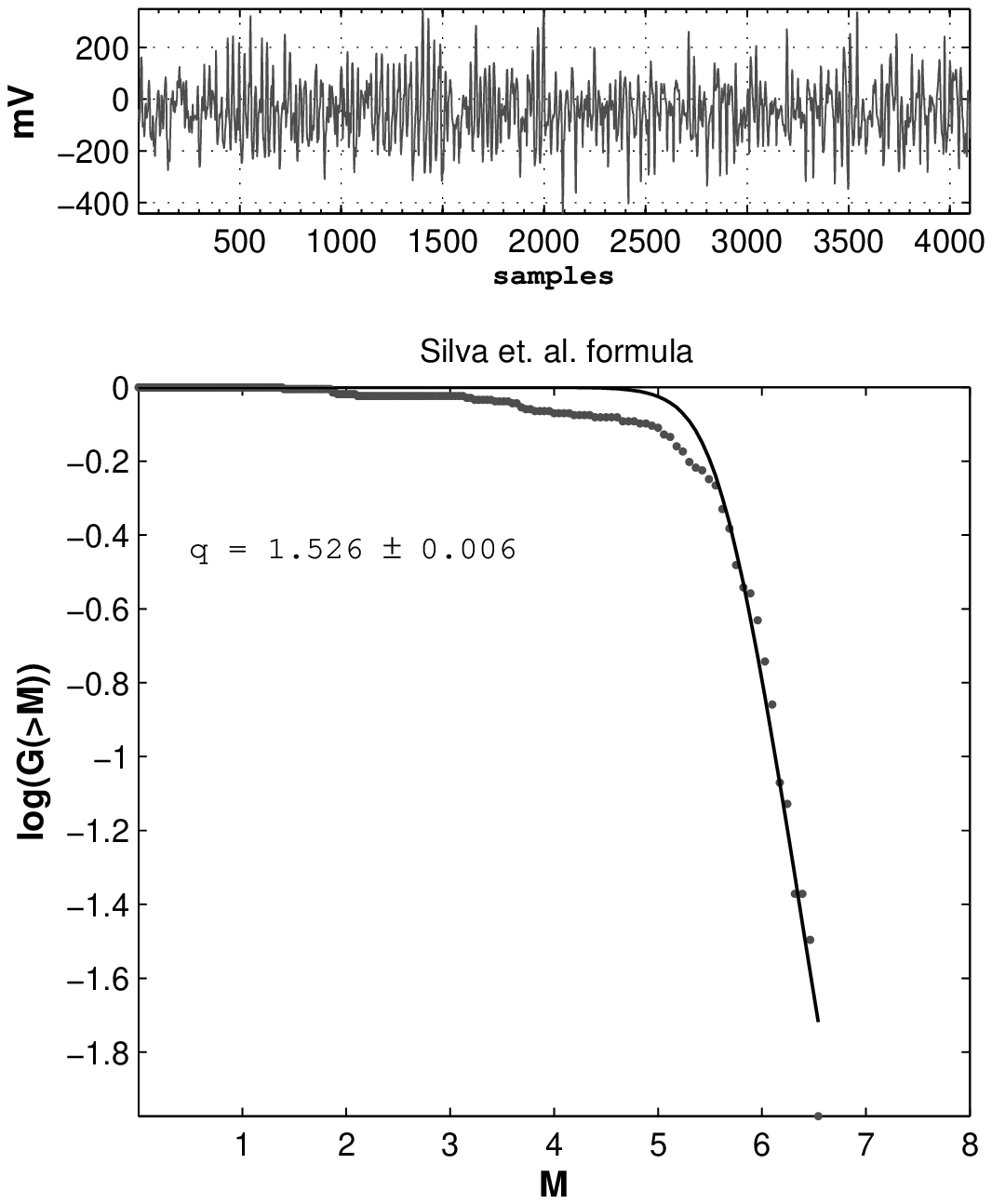}}
\end{center}
\caption{We used Eq. \ref{eq:silva} to fit the distribution of the magnitudes of electric events included in (a) the sequence of 100 seizures belonging to the ``E'' set \cite{Andrzejak2001}, (b) a single human seizure included in the set ``E''.}
\end{figure*}

Fig. \ref{subfig:silva_100EEGs} shows that the distribution of magnitudes of electric pulses included in the 100 seizures of the ``E'' set \cite{Andrzejak2001} is also described by the formula (\ref{eq:silva}). The best-fit parameter for this analysis is given by $q=1.573$. Figs. \ref{subfig:silva_1EEG} shows the same application for a single human seizure ($q=1.526$), correspondingly.

Though intriguing to some extent, the above mentioned results reveal that the obtained formula (\ref{eq:silva}) is not a mere artefact, and suggests that a more exhaustive study of the aforementioned biological and geophysical shocks in terms of nonextensive statistics is needed to give a deeper interpretation of their generation.

In summary, the existence of dynamical analogy between earthquake dynamics and neurodynamics has been supported by the analysis by means the of nonextensive Eq. (\ref{eq:silva}).

\section{Dynamical analogy in terms of waiting times}

Power-law correlations in both space and time are at least required in order to verify dynamical analogies between different catastrophic events. Hence, one can ask how the EM fluctuations included in a single EM precursor and electrical fluctuations included in a single epileptic seizure correlate in time. We investigate the aforementioned temporal clustering in terms of burst lifetime (duration) and waiting time $\tau$ (time interval between two successive electric events) focusing on a potential power-law distribution.

In \cite{Osorio2010}, the probability-density-function for intervened intervals $\tau$ were calculated for a population of different seizures and earthquakes. Both statistics follow a power-law distribution $\sim 1/(\tau^{1+\beta})$, however, with different slopes, namely, ($\beta\sim 0.1$) for earthquakes and $\beta\sim 0.5$ for interseizure intervals.

\begin{figure*}[ht]
\begin{center}
\subfloat []{\label{subfig:EM_wtimes}\includegraphics[width=0.45\textwidth]{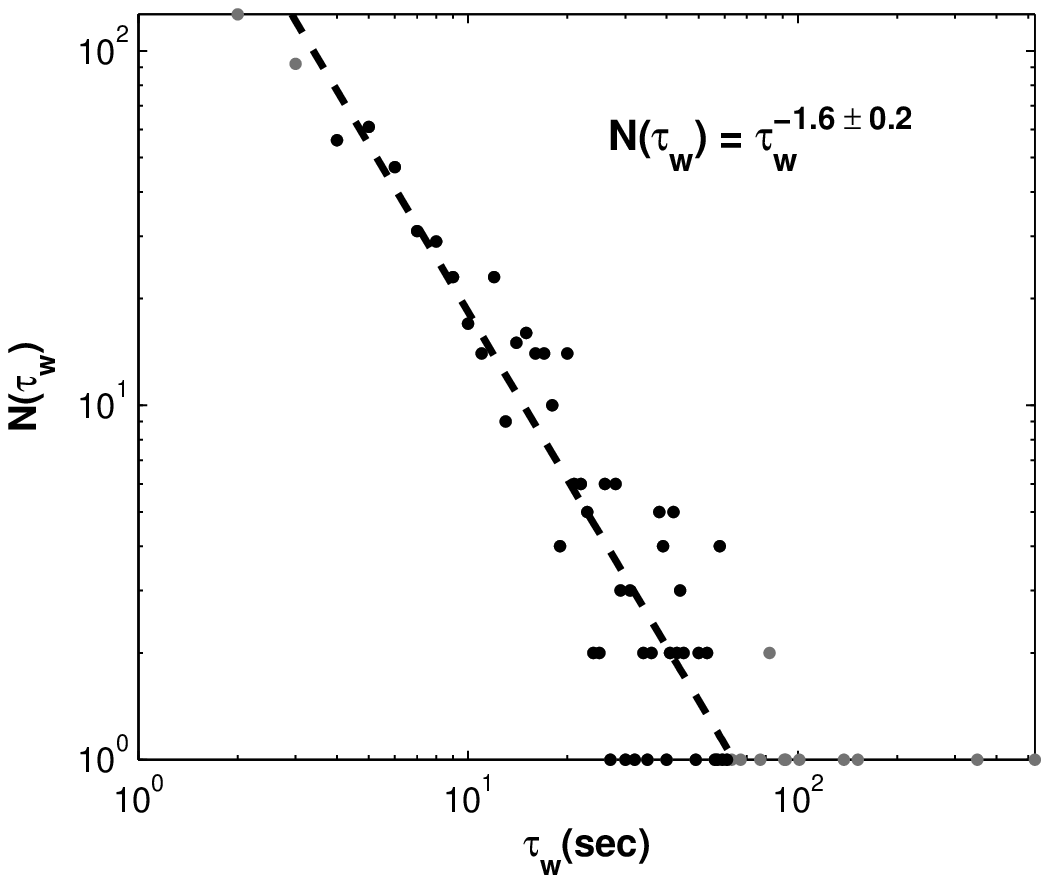}}
\subfloat []{\label{subfig:EM_dtimes}\includegraphics[width=0.45\textwidth]{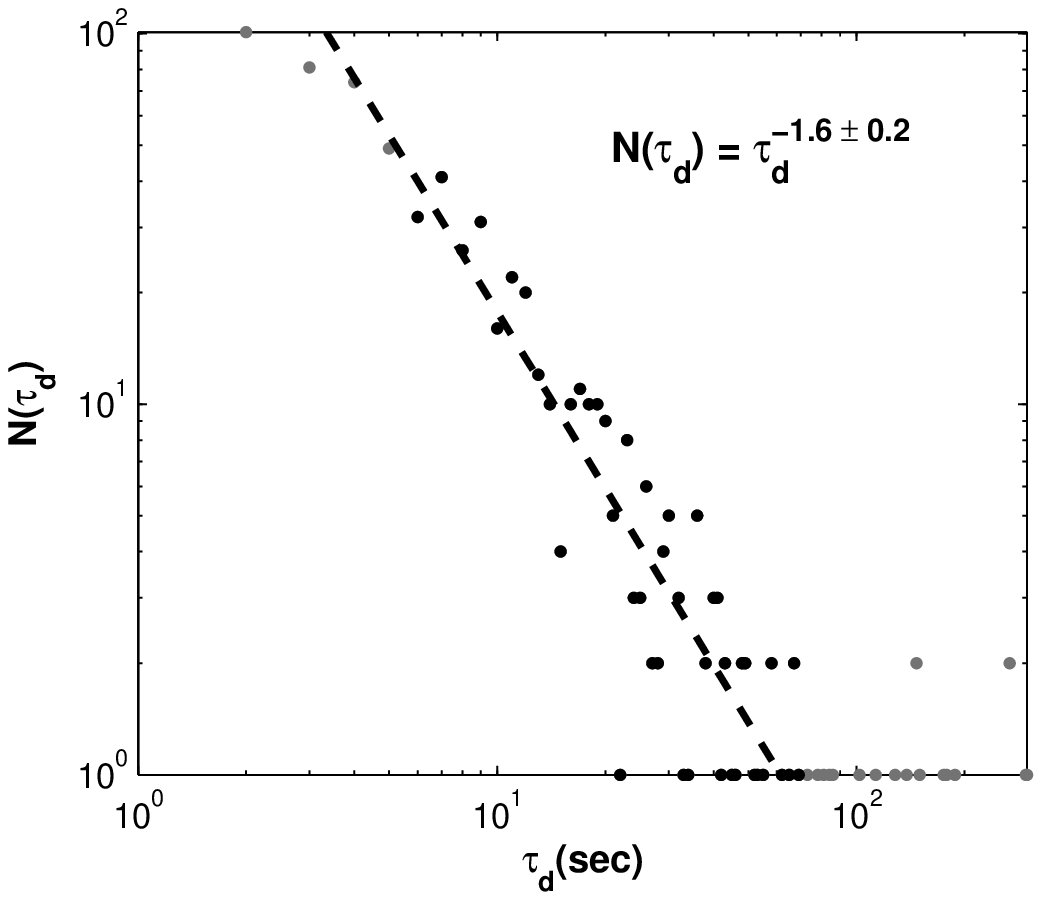}}
\end{center}
\caption{The figure refers to the precursory kHz EM activity associated to Athens earthquake. (a) The waiting times until the next EM fluctuation display the power-law distribution $\sim 1 / \tau^{1.6}$. (b) The time durations (livetime) of the emerged EM events, follows the power-law distribution $\sim 1 / \tau^{1.6}$.}
\end{figure*}

We refer to the pre-seismic EM activity associated with the Athens earthquake (see Fig. \ref{fig:EM_entropies} upper panel). The analysis reveals that the waiting times until the next EM fluctuation display a power-law distribution $\sim 1 / {\tau_{w}}^{1.6}$ (see Fig. \ref{subfig:EM_wtimes}). The same power-law is followed by the distribution of lifetimes (Fig. \ref{subfig:EM_dtimes}). We note that Vespignani et al. \cite{Vespignani1995} measured a corresponding exponent $\sim 1.6$ via acoustic signals from laboratory samples subjected to an external stress. We observe a similarity of critical exponents associated with the description of the population of: (i) EM fluctuations included in a single EM precursor rooted in the activation of a individual fault, and (ii) different seizures \cite{Osorio2010}. This finding further enhances the proposal that a dynamical analogy exists between fracture phenomena and seizures.

\section{The role of the coupling strength (or heterogeneity)}

Osorio et al. \cite{Osorio2010}, have examined the role of the coupling strength (or heterogeneity) in the dynamical behaviour of the excited brain. As shown in the generic phase diagram of Fiq. 6 of their work, for very weak coupling and large heterogeneity, the dynamics are incoherent; increasing the coupling strength (and/or decreasing the heterogeneity) leads to the emergence of intermediate coherence and of a power-law critical regime. Further increase in coupling strength (and / or decreases in heterogeneity) force the system towards strong coherence or synchronization and periodic behaviour. The generic phase diagram leads to the prediction that, if the degree of the coupling strength (or of the heterogeneity) between threshold oscillators is manipulated, transitions between the criticality and synchronized regimes will not only occur but will be coextensive.

\begin{figure*}[ht]
\begin{center}
\subfloat []{\label{subfig:rat_silva}\includegraphics[width=0.33\textwidth]{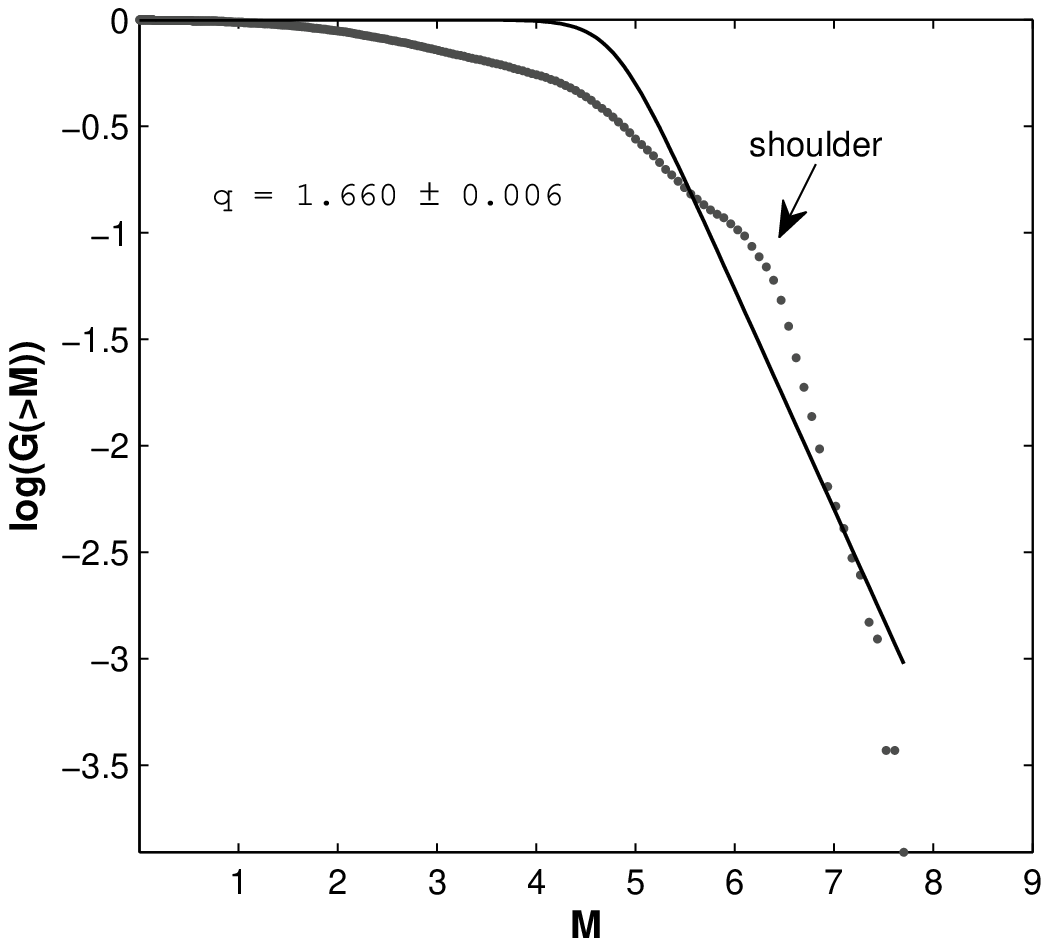}}
\subfloat []{\label{subfig:rat_wtimes}\includegraphics[width=0.33\textwidth]{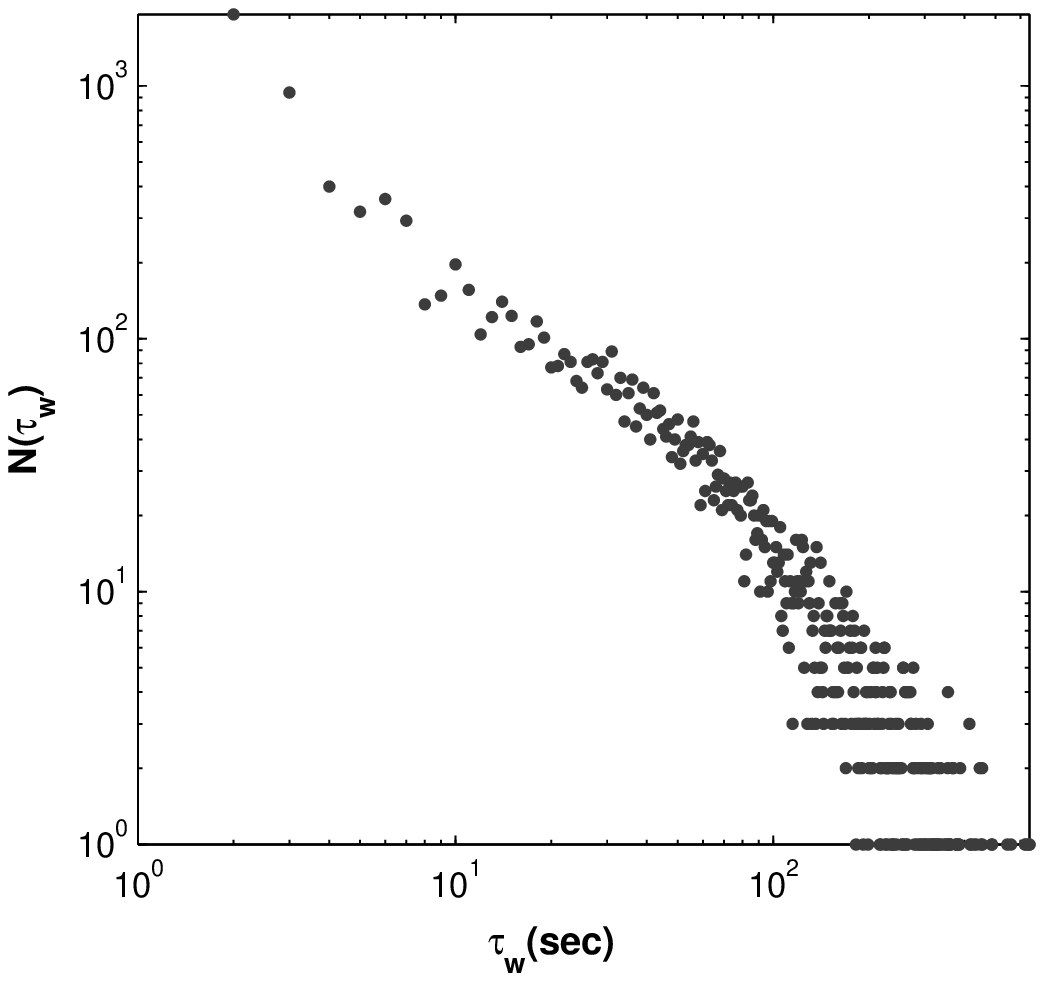}}
\subfloat []{\label{subfig:rat_dtimes}\includegraphics[width=0.33\textwidth]{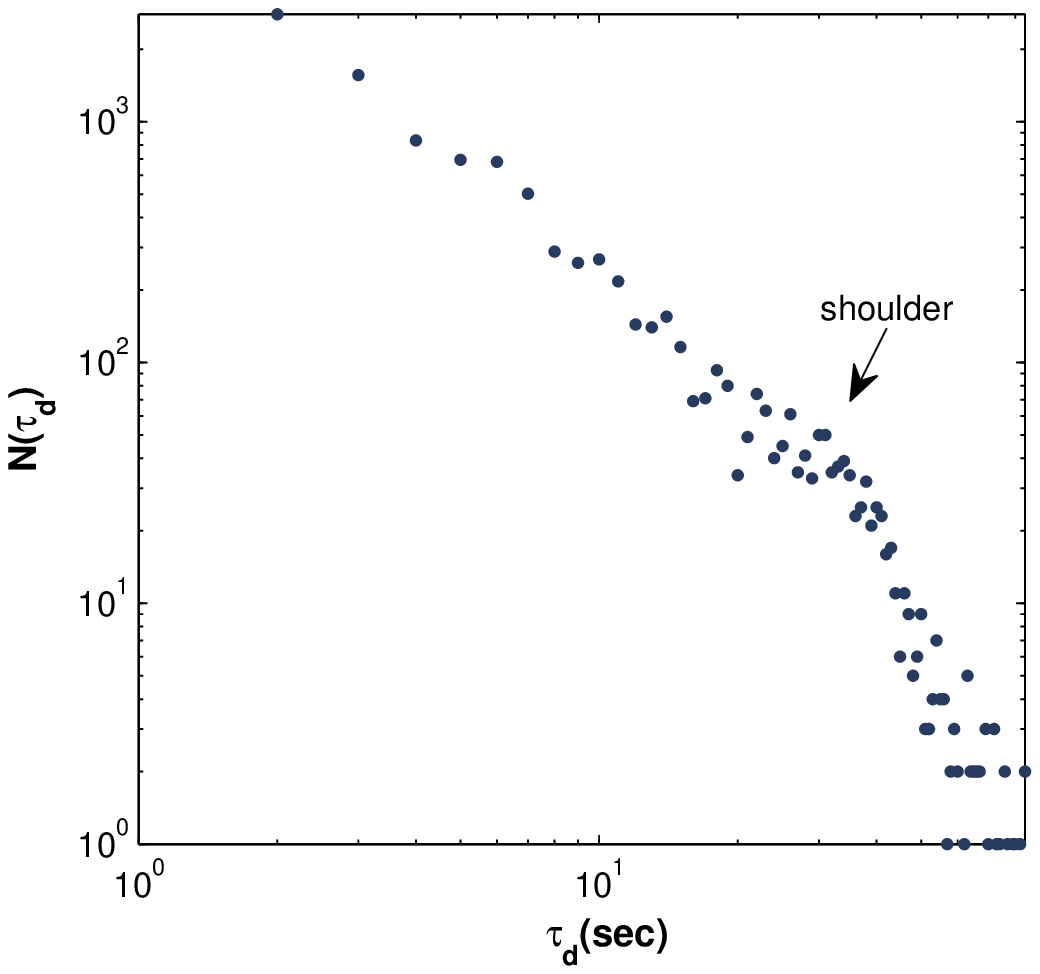}}
\end{center}
\caption{The distributions of the magnitudes (a), of the electric pulses included in the rat seizure of Fig 1 (b) waiting times (c), and life-times. The arrow indicates the characteristic ``shoulder''(see text). This evidence indicates that transitions between the criticality and synchronized regimes have been occurred \cite{Osorio2010}.}
\end{figure*}

The rat seizure shown in Fig. \ref{fig:EEG_entropies} presents the above mentioned situation. Indeed, Fig. \ref{subfig:rat_silva} depicts the variation of $\log G(>M)=N(M>)/N$ vs $M$, where $G(>M)=N(M>)/N$ is the relative cumulative number of electric pulses included in this single seizure with magnitude larger than $M$. We observe that the epileptic electric pulses with magnitude around 6 violate the nonextensive formula (\ref{eq:silva}) forming a characteristic ``shoulder''. This finding is inductive of the existence of a ``characteristic'' size (magnitude) in the distribution of $M$. We underline that this characteristic ``shoulder'' has been also appeared in the distribution of seizure energies of rats (see Fig. 7 in \cite{Osorio2010}). Fig. \ref{subfig:rat_wtimes} refers to the distribution of the waiting times of the emerged electric pulses is depicted. We observe that the data seem to form a ``shoulder'', as it is happened in the distribution of the magnitudes (Fig. \ref{subfig:rat_silva}). This formation of a ``shoulder'' is more clear in Fig. \ref{subfig:rat_dtimes}, where the distribution of lifetime (duration) of the emerged epileptic electric pulses is depicted.

The above mentioned results verify the proposal that transitions between the criticality and synchronized regimes will not only occur but could be coextensive.

\section{Discussion \& Conclusions}

Epileptic seizures and pre-seismic electromagnetic (EM) emissions are characteristic complex phenomena. They are complex in the sense that there are a great many apparently independent agents (firing neurons / opening cracks) interacting with each other and it is the richness of these interactions that allows the system as a whole to undergo self-organization. Another key feature of these phenomena is non-linearity and feedback loops in which small changes can have striking effects that cannot be understood simply by analysing the individual components. That is, the whole is more than the sum of its (reductionist) parts \cite{Pearce2006}.

The field of study of complex systems holds that the dynamics of complex systems are founded on universal principles that may be used to describe disparate problems. Complexity theory is influencing fields as diverse as physics, cosmology, chemistry, geography, climate research, zoology, biology, evolutionary biology, cell biology, neuroscience, clinical medicine, management, and economics \cite{Pearce2006}. The strong analogies between earthquake dynamics and neuro-dynamics have been drawn by numerous authors suggesting the applicability of analysis within similar mathematical frameworks. In this frame, the main objective of the present work was to examine whether a unified theory could really exist for the ways in which firing neurons / opening cracks organize themselves to produce a single epileptic seizure / earthquake.

The main contribution of this paper is to suggest a shift in emphasis from ``large'' to ``small'', in terms of scale, in the search for a dynamical analogy between seizure and earthquake. Our analyses examine a single epileptic seizure generation and the activation of a single fault (earthquake) and not the statistics of sequences of different seizures and earthquakes, as in previous studies \cite{Sornette2002}. A central property of the epileptic seizure / earthquake generation is the occurrence of coherent large-scale collective behaviour with very rich structure, resulting from repeated nonlinear interactions among the constituents of the system, respectively firing neurons and opening cracks. Consequently, the Tsallis nonextensive statistical mechanics has been found to be an appropriate framework in order to investigate dynamical analogies between epileptic and seismic crises. For completeness reasons we have also used entropic measures as well as tools from information theory.

The analytical approach employed in this paper has unfolded in two stages. First, we examined the data in terms of multidisciplinary statistical analysis methods, aiming to discover common ``pathological'' symptoms of transition to a significant seismic or epileptic shock. By monitoring the temporal evolution of the degree of organization in EEG rat and human time series and pre-seismic EM time series associated with the activation of a single fault (earthquake), we observed similar distinctive features revealing significant reduction of complexity during their emergence. These alternations were shown with Tsallis entropy and further enhanced a by T-entropy, approximate entropy, and Shannon block entropies and Fisher information. Interestingly, the transition from anti-persistent to persistent behaviour indicates the onset of the two crises under study. We note that the same pathological symptoms characterise the emergence of magnetic storms \cite{Balasis2006,Balasis2008,Balasis2009a,Balasis2011}. Secondly, we examined the existence of dynamical correspondences between a single seizure and a single earthquake generation by means of scale-free statistics, namely, a common hierarchical organization that results in power-law behaviour over a significant range of values of some parameters such as event energy or waiting time. We concentrate on the question whether the corresponding power-laws, if any, share the same exponent. Examining the data in terms of conventional and nonextensive dynamic scale-free laws we found that the magnitudes / energies and waiting times of the included EM pulses / electric pulses in a single EM precursor / single seizure follow a power-law distribution with a rather similar exponent. This finding is extended to the laboratory seismicity by means of acoustic emission, as well. Differences in constituting elements (organic vs inorganic), in scale and other physical properties between the earth and brain may account for small dissimilarities in exponents of the statistics \cite{Osorio2010}. These findings enhance the existence of dynamical analogies between the two complex phenomena under study. Importantly, similar power-law distributions are followed by sequences of different natural earthquakes and different seizures. This result supports the self-affine nature of earthquake / seizure generation.

It has been proposed that if the degree of the coupling strength (or of the heterogeneity) between threshold oscillators is manipulated, transitions between the criticality and synchronized regimes will not only occur but will be coextensive. This feature has also been recognized by means of nonextensive and  conventional scaling laws in the case of rat epilepsy.

A promising concept developed early on \cite{Keilis-Borok1964,Mogi1985} and extended in the last decade views a large earthquake as the culmination of a preparatory phase during which smaller earthquakes smooth out the stress field and express the long-range correlation of stresses that could be associated with the large runaway \cite{Kossobokov2002,Keilis-Borok2004}. This corresponds to viewing a large earthquake as a kind of dynamical critical point \cite{Bowman1998,Jaume1999}, in which accelerated seismic release results from a positive feedback of the seismic activity on its release rate \cite{Sammis2002}. Accumulated evidence support that seizures can be also viewed as dynamical critical points; the approach of ``intermittent criticality'' offers a possible common scenario for the development of severe epileptic and seismic shocks \cite{Eftaxias2006}. The generation of magnetic storms also seems to follow the aforementioned approach \cite{Balasis2006} .

Future tests of the seizure-earthquake analogy should also involve the question of seismic localization (faults) versus seizure focus/epileptogenic zone as conventionally defined \cite{Osorio2010}. There are already arguments implying that the main seismic / epileptic shock occurrence is accompanied by the appearance of a preferred direction of elementary activities \cite{Eftaxias2006}.

The study of EM precursory activity refers to earthquakes that have occurred in land or near coast-line, they are shallow and have a magnitude of approximately 6 or higher \cite{Eftaxias2002,Eftaxias2004,Eftaxias2006,Kapiris2004,Karamanos2006}. However, EQs that meet all these conditions do not occur frequently. Thus, there is an inherent limitation in this research field, in terms of the amount of data (precursory EM recordings from different EQs) that can be used for the investigation of patterns and distinctive features of preseismic EM activity. The case is quite different in the study of EEGs related to epileptic seizures, as: (i) it is known a priori that data recorded relates to a diagnosed epileptic case, (ii) there is a surplus in terms of quantity and diversity of data available for analysis, and subsequently (iii) it is possible to perform controlled laboratory analysis.

On these grounds, this paper comes to propose that as long as similar properties and common distinctive features can be found in the two types of extreme events, it may be possible to draw on identified dynamic analogies in order to utilize transferable ideas and methods. Results and insights from the study of epileptic seizures may provide useful approaches that can feed back into the analysis of preseismic EM activity and enhance our ability to identify potential precursory EM patterns.

In summary, the obtained results support the claim that epileptic seizures might be considered as ``quakes of the brain''.

\section*{Acknowledgments}
The second author (G.M) would like to acknowledge research funding received by the Greek State Scholarships Foundation (IKY).


\end{document}